\documentclass[usenatbib]{mn2e}
\usepackage{epsfig}
\usepackage{amsmath}

\DeclareGraphicsExtensions{.pdf,.ps,.eps,.png,.jpg,.mps} 
\usepackage{float,latexsym,graphicx,footnote}%,margins}%,ctable,pdflscape,float,subfig}
\usepackage{color}

\usepackage{times} 
\usepackage{url}
\newcommand{\be}{\begin{equation}} 
\newcommand{\ee}{\end{equation}} 
\newcommand{\bed}{\begin{displaymath}} 
\newcommand{\eed}{\end{displaymath}} 
\newcommand{\ba}{\begin{eqnarray}} 
\newcommand{\ea}{\end{eqnarray}}

\newcommand{\ww}{$\langle w\rangle$} 

\def\ls{\mathrel{\hbox{\rlap{\hbox{\lower4pt\hbox{$\sim$}}}\hbox{$<$}}}}
\def\gs{\mathrel{\hbox{\rlap{\hbox{\lower4pt\hbox{$\sim$}}}\hbox{$>$}}}}
\def\wwm{\mathrel{\langle w\rangle}}
\def\Mx{\mathrel{M_{\rm X}}}
\def\Mwl{\mathrel{M_{\rm WL}}}
\def\Ysph{\mathrel{Y_{\rm sph}}}
\def\Lx{\mathrel{L_{\rm X}}}
\def\Yx{\mathrel{Y_{\rm X}}}

\begin{document}

\title[LoCuSS Hydrostatic Masses]{LoCuSS: Hydrostatic Mass
    Measurements of the High-$\Lx$ Cluster Sample -- Cross-calibration
    of \emph{Chandra} and \emph{XMM-Newton}}

\author[Martino et al.]{Rossella Martino$^{1,2}$, Pasquale Mazzotta$^{1}$, Herv\'{e} Bourdin$^{1}$, Graham P.\ Smith$^{3}$,
\newauthor  Iacopo Bartalucci$^{1}$, Daniel P.\ Marrone$^{4}$,  Alexis Finoguenov$^{5}$, Nobuhiro Okabe$^{6}$\\
$^{1}$Dipartimento di Fisica, Universit\`a degli Studi di Roma ``Tor Vergata'', via della Ricerca Scientifica 1, 00133, Roma,  Italy\\
$^{2}$Laboratoire AIM, IRFU/Service dÕAstrophysique -CEA - CNRS, B‰t. 709, CEA-Saclay, 91191 Gif-sur-Yvette Cedex, France\\
$^{3}$School of Physics and Astronomy, University of Birmingham, Birmingham, B15 2TT, England\\
$^{4}$Steward Observatory, University of Arizona, 933 North  Cherry Avenue, Tucson, AZ 85721, USA\\
$^{5}$Department of Physics, University of Helsinki, Gustaf H\"allstr\"omin katu 2a, FI-00014 Finland\\
$^{6}$Academia Sinica Institute of Astronomy and Astrophysics (ASIAA), P.O. Box 23-141, Taipei 10617, Taiwan\\
}

\maketitle

%%%%%%%%%%%%%ABSTRACT%%%%%%%%%%%%%%%%%%%%%
\begin{abstract}
  We present a consistent analysis of \emph{Chandra} and
  \emph{XMM-Newton} observations of an approximately mass-selected
  sample of 50 galaxy clusters at $0.15<z<0.3$ -- the ``LoCuSS High-$L_X$
  Sample''.  We apply the same analysis methods to data from both
  satellites, including newly developed analytic background models
  that predict the spatial variation of the \emph{Chandra} and
  \emph{XMM-Newton} backgrounds to $<2\%$ and $<5\%$ precision
  respectively.  To verify the cross-calibration of \emph{Chandra}-
  and \emph{XMM-Newton}-based cluster mass measurements, we derive the
  mass profiles of the 21 clusters that have been observed with both
  satellites, extracting surface brightness and temperature profiles
  from identical regions of the respective datasets.  We obtain
  consistent results for the gas and total hydrostatic cluster masses:
  the average ratio of \emph{Chandra}- to \emph{XMM-Newton}-based
  measurements of $M_{\rm gas}$ and $M_X$ at $r_{500}$ are
  $0.99\pm0.02$ and $1.02\pm0.05$, respectively with an intrinsic
  scatter of $\sim3\%$ for gas masses and $\sim8\%$ for hydrostatic
  masses.  Comparison of our hydrostatic mass measurements at
  $r_{500}$ with the latest LoCuSS weak-lensing results
  indicate that the data are consistent with non-thermal
  pressure support at this radius of $\sim7\%$.  We also investigate
  the scaling relation between our hydrostatic cluster masses and
  published integrated Compton parameter $\Ysph$ measurements from the
  Sunyaev-Zel'dovich Array.  We measure a scatter in mass at fixed
  $\Ysph$ of $\sim16\%$ at $\Delta=500$, which is consistent with
  theoretical predictions of $\sim10-15\%$ scatter.
\end{abstract}


%%%%%%%%%%%%%%%INTRODUCTION%%%%%%%%%%%%%%%%%%%

\section{Introduction}
\par
Galaxy clusters are the most massive gravitationally collapsed objects
in the universe. Their abundance is determined by the spectrum of
primordial density perturbations, which are amplified by gravity over
the expansion history of the universe.  They are therefore intricately
connected to many cosmological parameters and may be used to determine
precise values for them \citep{1974ApJ...187..425P,
  1993Natur.366..429W,1998MNRAS.298.1145E, 2002MNRAS.334L..11A,
  2009ApJ...692.1060V}.

Through numerical simulations it is possible to reproduce the cluster
formation mechanism, to predict the detailed shape of the mass
function \citep{1999MNRAS.308..119S, 2001MNRAS.321..372J,
  2008MNRAS.386.2135G, 2012ApJ...745...16T} and to identify expected
structural and scaling properties \citep[e.g][]{1997ApJ...490..493N,
  1986MNRAS.222..323K, 2006ApJ...650..128K, 2005Natur.435..629S}.
These simulations provide the basis for the the cosmological utility
of galaxy clusters, but their detailed predictions must be examined
against high-quality observations in order to achieve the desired
cosmological precision.  Such tests can indicate the need for improved
physics in the modeling, identify biases in the observational
procedures
\citep{2006MNRAS.369.2013R,2010A&A...514A..93M,2007ApJ...655...98N},
and characterize the intrinsic uncertainties in the mass observables
being employed.  In particular, intercomparison of multiple mass
estimators can provide additional checks on our understanding
\citep[e.g.,][]{2010ApJ...711.1033Z,2013ApJ...767..116M}.

Cluster masses may be estimated from observations of the X-ray
emission produced by the hot intracluster medium (ICM) within galaxy
clusters.  In particular, cluster mass profiles can be derived from
the gas density and temperature structure, assuming the hot ICM to be
in hydrostatic equilibrium within the cluster gravitational potential
wells.  With respect to galaxy dynamics or lensing mass estimates,
this method has the advantage of being less sensitive to projection
effects due to mass along the line of sight through the cluster.
However, validity of the assumptions of ICM hydrostatic equilibrium
and spherical symmetry of the cluster gravitational potential wells
may depend on the evolutionary state of the cluster.

The launch of the {\it Chandra} and {\it XMM-Newton} satellites has
greatly improved the resolution, sensitivity, and precision of X-ray
observations of galaxy clusters.  Both satellites can identify clusters
well beyond $z=1$, discerning morphology and permitting temperature
and mass measurements to similar distance
\citep[e.g.,][]{2008MNRAS.387..998M, 2012ApJ...761..183S, 2012AA...539A.105S}.

However, there remain inconsistencies between measurements with the
two observatories \citep[e.g.,][]{2006ApJ...640..691V,
  2008A&A...478..615S,2010A&A...523A..22N,2011A&A...525A..25T,2013ApJ...767..116M}. The
difference between the point spread function (PSF) of the two
instruments as well as different methods of background subtraction
often make direct comparison difficult. The background treatment is
particularly important far from cluster centers where the most
cosmologically useful masses are measured. Clusters fade rapidly into
the background at large radii, so the reliability of the background
modeling determines the radial range over which cluster parameters can
be accurately determined.

In this work, we describe  X-ray data analysis technique for {\it
  Chandra} and {\it XMM-Newton} observations that is intended to
provide consistent hydrostatic masses from either satellite at
cosmologically interesting radii.  Crucially, for both {\it
  Chandra} and {\it XMM-Newton}, we model analytically the spatial variations of all background noise components  \citep{Bart2014}.
We apply this technique to the 50 galaxy clusters of the LoCuSS
High-$\Lx$ sample (\S\ref{sample}), estimating hydrostatic
masses from all available observations from both satellites.  From the
sub-sample of 21 clusters observed by both, we characterize the
consistency of the masses obtained from the two satellites under a
consistent analysis procedure.  Relationships between the X-ray masses
obtained from our procedures and other cluster observables are also
examined.

We describe the sample in \S\ref{sample}.  In
\S\S\ref{xray_an},\ref{profiles},\ref{sec:morph} we describe the
details of the data analysis and the method of mass estimation,
emphasizing the novel aspects of our technique. In \S\ref{results}, we
examine the consistency of the mass, density, and temperature profiles
between satellites. We compare our X-ray mass estimates with recent
weak-lensing mass estimates to investigate observational
biases. Finally, we characterize the relationship between the X-ray
masses and the Sunyaev-Zel'dovich (SZ) effect signature of clusters
($Y_{sz}$) and quantify its scatter, comparing our results with
simulations and previous works
\citep{2006ApJ...650..538N,2012ApJ...754..119M,2011ApJ...738...48A}.
We adopt a $\Lambda$CDM cosmology: $\Omega_M=0.30$,
$\Omega_{\Lambda}=0.70$, $H_0=70 \rm{km\, s^{-1} Mpc^{-1}}$.

%%%%%%%%%%%%%%%%%%%%DATA ANALYSIS%%%%%%%%%%%%%%%
\section{Sample and Observations}
\label{sample}

\begin{figure}

\includegraphics[width=.37\textwidth,angle=-90]{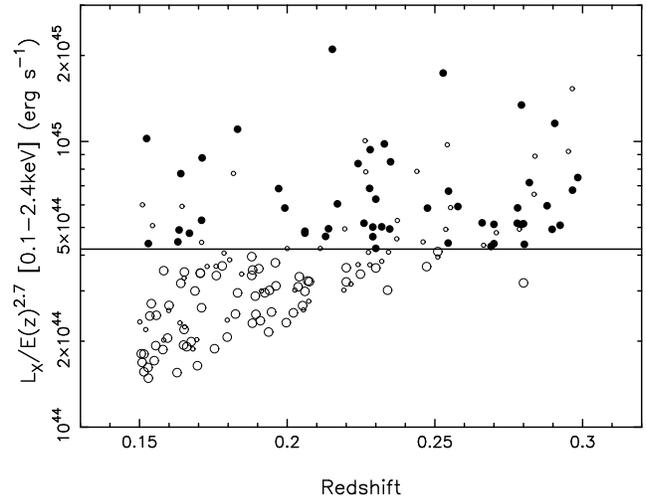}
\caption{The LoCuSS High-$\Lx$ Sample, as defined in \S2 (filled large
  circles).  Large open circles identify clusters that satisfy the
  declination cut for the High-$\Lx$ sample but not the luminosity cut
  (horizontal line).  Small open circles do not satisfy the
  declination cut.  Clusters from the RASS parent catalogues are only
  plotted if they satisfy the initial selection described in \S2.  The
  absence of points from the bottom right corner of the figure is due
  to the RASS flux limit.}
\label{Locuss_sample}
\end{figure}

We study a sample of 50 clusters selected from the \emph{ROSAT} All
Sky Survey catalogs \citep[RASS;][]{2000MNRAS.318..333E,
  2004A&A...425..367B}.  These clusters are the ``High-$\Lx$'' sample
defined by the Local Cluster Substructure Survey
(LoCuSS\footnote{\url{http://www.sr.bham.ac.uk/locuss}}).  The initial
LoCuSS selection from the RASS catalogs applied the following
criteria: $0.15<z<0.3$, $n_H<7\times10^{20}{\rm cm}^{-2}$,
$-70^\circ<\delta<70^\circ$ to obtain a sample of 165 clusters that
span X-ray luminosities of $2\times10^{44}\ls\Lx[0.1-2.4{\rm
    keV}]\ls2\times10^{45}{\rm erg\,s^{-1}}$.  The 50 clusters studied
in this article satisfy the following additional criteria:
$-25^\circ<\delta<65^\circ$, $\Lx[0.1-2.4{\rm
    keV}]\,E(z)^{-2.7}\ge4.2\times10^{44}{\rm erg\,s^{-1}}$.  The
first criterion ensures that the clusters are observable at high
elevation from the Subaru 8-m telescope on Mauna Kea, and the second
implements an \emph{approximate} mass selection, by adopting the
\cite{2005A&A...433..431P} mass-$\Lx$ scaling relation:
$\Lx\,E(z)^{-2.7}\propto M$, where $E(z)\equiv H(z)/H_0$ is the
evolution of the Hubble parameter, and ignoring scatter in the
mass-$\Lx$ relation (Table~\ref{tab:cluster sample}; Figure~\ref{Locuss_sample}).  This article presents a new
X-ray analysis of the entire High-$\Lx$ sample.  The sub-sets of
our sample of 50 clusters studied by Okabe et al.\ (2010), Marrone et
al.\ (2012), and Mahdavi et al.\ (2012) are identified in Table~\ref{tab:cluster sample}.

All 50 clusters have been observed with either or both of
\emph{Chandra} and \emph{XMM-Newton} (Table~1).  The \emph{Chandra}
and \emph{XMM-Newton} data are obtained with the Advanced CCD Imaging
Spectrometer (ACIS) camera and the European Photon Imaging Camera
(EPIC) respectively.  The \emph{Chandra} ACIS camera contains 10
planar 1024 x 1024 pixels CCDs: four arranged in a 2x2 array (ACIS-I),
and six arranged in a 1x6 array (ACIS-S)\protect\footnote{For more details see
  \url{http://cxc.cfa.harvard.edu/proposer/POG/html/ chap6.html}}.
The \emph{XMM-Newton} EPIC is the combination of three cameras: two
MOS and a PN\protect\footnote{For more details see
  \url{http://xmm.esac.esa.int/external/xmm_user_support/
    documentation/technical/EPIC/} }.

From the full sample of 50 clusters, 43 have been observed with
\emph{Chandra}, and 39 have been observed with \emph{XMM-Newton}.
After excluding observations strongly affected by flares
(\S\S3.1~\&~3.2), useful data are available for all 50 clusters
  from either/both satellites.  Importantly, useful data are available
  from \emph{both} satellites for 27 clusters; the \emph{Chandra}
data for 21
of these 27 were obtained with ACIS-I.  This relatively large sample
of clusters with data from both satellites allows us to quantify the
uncertainties on the hydrostatic mass measurements that are caused by
instrumental cross-calibration issues.  This is a main focus of our
study (\S6).

\noindent\begin{table*}
  \caption{Cluster Sample and X-ray Observations.\protect\\ 
  ${^a}$Low and high centroid shift categories are
      defined as \protect$\langle w\rangle<0.01r_{500}$ and \protect$\langle w\rangle>0.01r_{500}$, respectively.\protect\\
   ${^b}$Sub-samples defined thus: (1) observed with
      \emph{XMM} and \emph{Chandra} (ACIS-I VF Mode), (2)
    included in the \protect\cite{2012ApJ...754..119M} sample, (3)
      included in the \protect\cite{2010PASJ...62..811O} sample, (4)
      included in the \protect\cite{2013ApJ...767..116M} sample.\protect\\
      ${^c}$\emph{Chandra} ACIS-S data.\protect\\
    ${^d}$Data affected by flaring episodes for $>90\%$ of the total exposure time; data not used for mass estimates.}
  \resizebox*{0.7\textheight}{!}{
      \begin{tabular}{p{24mm}p{7mm}p{11mm}p{14mm}p{20mm}p{33mm}p{15mm}p{17mm}p{5mm}p{5mm}p{8mm}}
        \hline
        \noalign{\smallskip}
        Cluster  & $z$ & \multispan2{\hfil $\alpha,\delta$\,[J2000]\hfil} & \multispan2{~Observation IDs\dotfill}  & \multispan2{\hfil Centroid shift, $\langle w\rangle/(0.01r_{500})$\hfil} & \multispan2{~Sub-samples$^{a,b}$} & $N_{\rm{Bin,T}}$  \\
                             &         &             &               & \emph{Chandra} & \emph{XMM-Newton} & \emph{Chandra} & \emph{XMM-Newton} &  & \\
        \noalign{\smallskip}
        \hline
        \noalign{\smallskip}
        ABELL\,2697          & $0.232$ & 00:03:11.98 & $-$06:05:35.6 & x         & 0652010401     & x             & $0.11\pm0.02$ & Low   &   &5 \\
        \noalign{\smallskip}
        ABELL\,0068          & $0.255$ & 00:37:06.11 & $+$09:09:30.4 & 3250      & 0084230201     & $1.10\pm0.13$ & $1.05\pm0.05$ & High & 1,2,3,4 & 2 \\
        \noalign{\smallskip}
        ABELL\,2813          & $0.292$ & 00:43:24.72 & $-$20:37:24.8 & 9409      & 0042340201     & $1.25\pm0.09$ & $1.06\pm0.04$ & High   & 1  &2 \\
        \noalign{\smallskip}
        ABELL\,0115s         & $0.197$ & 00:56:00.30 & $+$26:20:32.5 & 3233      & 0203220101     & $5.52\pm0.06$ & $5.06\pm0.04$ & High   & 1,3,4 & 3  \\
        \noalign{\smallskip}
        ABELL\,0141          & $0.230$ & 01:05:35.82 & $-$24:40:51.9 & 9410      & x              & $4.89\pm0.20$ & x             & High &   &2\\
        \noalign{\smallskip}
        ZwCl\,0104.4$+$0048  & $0.254$ & 01:06:49.53 & $+$01:03:22.0 & 10465$^c$ & x              & $0.34\pm0.04$ & x             & Low &  &2  \\
        \noalign{\smallskip}
        ABELL\,0209          & $0.206$ & 01:31:53.34 & $-$13:36:44.0 & 3579, 522 & 0084230301     & $0.70\pm0.10$ & $0.62\pm0.04$ & Low   & 1,2,3,4& 2 \\
        \noalign{\smallskip}
        ABELL\,0267          & $0.230$ & 01:52:42.32 & $+$01:00:40.9 & 3580      & 0084230401$^d$ & $2.23\pm0.05$ & x             & High   & 2,3,4 &2 \\
        \noalign{\smallskip}
        ABELL\,0291          & $0.196$ & 02:01:42.99 & $-$02:11:46.3 & x         & 0605000301     & x             & $0.46\pm0.04$ & Low   &  3 & 6 \\
        \noalign{\smallskip}
        ABELL\,0383          & $0.188$ & 02:48:03.38 & $-$03:31:45.6 & 2320, 524 & 0084230501     & $0.17\pm0.05$ & $0.17\pm0.04$ & Low   & 1,2,3,4 & 4 \\

        \noalign{\smallskip}
        ABELL\,0521          & $0.248$ & 04:54:08.23 & $-$10:14:23.6 & 430$^c$   & 0603890101     & $4.57\pm0.11$ & $5.28\pm0.04$ & High   & 2,3,4  & 2 \\
        \noalign{\smallskip}
        ABELL\,0586          & $0.171$ & 07:32:20.53 & $+$31:37:51.8 & 530, 11723 & 0605000801$^d$ & $0.26\pm0.07$ & x             & Low   & 2,3,4 & 4 \\
        \noalign{\smallskip}
        ABELL\,0611          & $0.288$ & 08:00:56.74 & $+$36:03:21.6 & 3194$^c$  & $0605000601^d$     & $0.33\pm0.08$ & x & Low   & 2,3,4  &2 \\
        \noalign{\smallskip}
        ABELL\,0697          & $0.282$ & 08:42:57.71 & $+$36:22:00.1 & 4217      & $0605000701^d$     & $0.49\pm0.08$ & x & Low   & 2,3,4 & 2 \\
        \noalign{\smallskip}
        ZwCl\,0857.9$+$2107  & $0.235$ & 09:00:36.88 & $+$20:53:41.5 & 10463$^c$ & 0402250701     & $0.18\pm0.08$ & $0.14\pm0.02$ & Low   & & 3 \\
        \noalign{\smallskip}
        ABELL\,0750          & $0.163$ & 09:09:12.67 & $+$10:58:26.6 & 924       & 0605000901$^d$ & $2.57\pm0.09$ & x             & High   & 3 &3 \\
        \noalign{\smallskip}
        ABELL\,0773          & $0.217$ & 09:17:52.67 & $+$51:43:38.1 & 533, 3588, 5066 & 0084230601     & $1.11\pm0.08$ & $1.39\pm0.04$ & High   & 1 & 3 \\
        \noalign{\smallskip}
        ABELL\,0781          & $0.298$ & 09:20:24.74 & $+$30:30:11.3 & 534       & 0150620201, 0401170101 & $5.43\pm0.25$ & $5.13\pm0.05$ & High   & 1 & 2 \\
        \noalign{\smallskip}
        ZwCl\,0949.6$+$5207  & $0.214$ & 09:52:49.15 & $+$51:53:04.5 & 3195$^c$  & x              & $0.33\pm0.08$ & x             & Low   &  & 2\\
        \noalign{\smallskip}
        ABELL\,0901          & $0.163$ & 09:55:57.24 & $-$09:59:04.7 & x         & 0148170101     & x             & $1.03\pm0.03$ & High   &  & 4  \\
        \noalign{\smallskip}
        ABELL\,0907          & $0.167$ & 09:58:22.18 & $-$11:03:51.6 & 3250, 3185, 535 & 0201903501, 0404910601& $0.30\pm0.04$ & $0.35\pm0.03$            & Low   &   1 & 6 \\
        \noalign{\smallskip}
        ABELL\,0963          & $0.205$ & 10:17:03.54 & $+$39:02:59.2 & 903$^c$   & 0084230701     & $0.25\pm0.03$ & $0.22\pm0.02$ & Low   &   4 & 5 \\
        \noalign{\smallskip}
        ZwCl\,1021.0$+$0426  & $0.291$ & 10:23:39.58 & $+$04:11:10.7 & 9371      & 0108670101     & $0.26\pm0.03$ & $0.39\pm0.02$ & Low   & 1 &2 \\
        \noalign{\smallskip}
        ABELL\,1423          & $0.213$ & 11:57:17.36 & $+$33:36:39.4 & 538, 11724 & x              & $1.27\pm0.15$ & x             & High   &  & 3   \\
        \noalign{\smallskip}
        ABELL\,1451          & $0.199$ & 12:03:17.25 & $-$21:32:14.0 & x         & 0652010101     &    x          & $0.76\pm0.05$ & Low   &  &5 \\
        \noalign{\smallskip}
        RXC\,J1212.3$-$1816  & $0.269$ & 12:12:18.10 & $-$18:18:31.4 & x         & 0652010201     &    x          & $2.66\pm0.12$ & High   & & 3   \\
        \noalign{\smallskip}
        ZwCl\,1231.4$+$1007  & $0.229$ & 12:34:21.97 & $+$09:47:10.0 & 539, 11727 & x              & $2.84\pm0.21$ & x             & High   &   & 2 \\
        \noalign{\smallskip}
        ABELL\,1682          & $0.226$ & 13:06:50.97 & $+$46:33:24.6 & 3244$^d$, 11725 & x              & $2.32\pm0.13$ & x             & High  &   & 2 \\
        \noalign{\smallskip}
        ABELL\,1689          & $0.183$ & 13:11:29.53 & $-$01:20:30.3 & 5004, 6930, 7289 & 0093030101     & $0.17\pm0.01$ & $0.23\pm0.02$ & Low   &1,4 & 6\\
        \noalign{\smallskip}
        ABELL\,1758          & $0.280$ & 13:32:43.74 & $+$50:32:45.0 & 2213$^d$  & 0142860201     &    x          & $1.45\pm0.03$ & High   &  4 & 5 \\
        \noalign{\smallskip}
        ABELL\,1763          & $0.228$ & 13:35:18.26 & $+$41:00:00.7 & 3591      & 0084230901     & $1.15\pm0.06$ & $1.54\pm0.07$ & Low   & 1,4 & 2 \\
        \noalign{\smallskip}
        ABELL\,1835          & $0.253$ & 14:01:01.92 & $+$02:52:42.8 & 6880, 6881, 7370 & \raggedright{0098010101, 0147330201, 0551830101} & $0.23\pm0.01$ & $0.37\pm0.01$ & Low   & 1,2,3,4 & 6\\
        \noalign{\smallskip}
        ABELL\,1914          & $0.171$ & 14:26:00.92 & $+$37:49:33.2 & 3593      & 0112230201     & $1.49\pm0.03$ & $1.13\pm0.01$ & High   & 1,3,4 & 3 \\
        \noalign{\smallskip}
        ZwCl\,1454.8$+$2233  & $0.258$ & 14:57:15.14 & $+$22:20:32.3 & 4192      & 0108670201     & $0.34\pm0.02$ & $0.29\pm0.02$ & Low   & 1,2,3 & 7 \\
        \noalign{\smallskip}
        ABELL\,2009          & $0.153$ & 15:00:19.50 & $+$21:22:10.3 & 10438     & x              & $0.17\pm0.03$ & x             & Low   &   &4  \\
        \noalign{\smallskip}
        ZwCl1459.4$+$4240    & $0.289$ & 15:01:20.84 & $+$42:20:57.4 & 7899      & 0402250201     & $1.36\pm0.16$ & $1.40\pm0.11$ & High   & 1,2,3 & 2\\
        \noalign{\smallskip}
        RXC\,J1504.1$-$0248  & $0.215$ & 15:04:07.51 & $-$02:48:15.2 & 5793      & 0401040101     & $0.12\pm0.01$ & $0.15\pm0.01$ & Low   & 1 & 5 \\
        \noalign{\smallskip}
        ABELL\,2111          & $0.229$ & 15:39:41.73 & $+$34:25:04.8 & 11726     & x              & $3.18\pm0.14$ & x             & High   & 4 & 2 \\
        \noalign{\smallskip}
        ABELL\,2204          & $0.152$ & 16:32:46.85 & $+$05:34:32.8 & 6104, 7940 & \raggedright{0306490201, 0306490301, 0306490401}     & $0.09\pm0.01$ & $0.22\pm0.02$ & Low   & 1,4 & 3 \\
        \noalign{\smallskip}
        ABELL\,2219          & $0.228$ & 16:40:21.20 & $+$46:42:18.4 & 896$^c$   & 0605000501     & $1.66\pm0.03$ & $1.31\pm0.04$ & High   & 2,3,4  & 8 \\
        \noalign{\smallskip}
        RX\,J1720.1$+$2638   & $0.164$ & 17:20:10.01 & $+$26:37:30.1 & 4361, 3224 & \raggedright{0500670201, 0500670301, 0500670401} & $0.26\pm0.05$ & $0.26\pm0.02$ & Low   & 1,3 & 5 \\
        \noalign{\smallskip}
        ABELL\,2261          & $0.224$ & 17:22:27.10 & $+$32:07:56.3 & 5007      & 0093030301$^d$ & $0.83\pm0.07$ & x             & Low   & 2,3,4 & 3 \\
        \noalign{\smallskip}
        RXC\,J2102.1$-$2431  & $0.188$ & 21:02:09.77 & $-$24:32:00.7 & x         & 0652010301     & x             & $0.34\pm0.04$ & Low   &  &5  \\ 
        \noalign{\smallskip}
        RX\,J2129.6$+$0005   & $0.235$ & 21:29:39.77 & $+$00:05:19.1 & 9370, 552 & 0093030201     & $0.42\pm0.11$ & $0.66\pm0.03$ & Low   & 1,2,3 & 2 \\
        \noalign{\smallskip}
        ABELL\,2390          & $0.232$ & 21:53:36.89 & $+$17:41:44.6 & 4193$^c$  & 0111270101     & $0.74\pm0.03$ & $0.64\pm0.03$ & Low   & 2,3,4 &9 \\
        \noalign{\smallskip}
        ABELL\,2485          & $0.247$ & 22:48:30.82 & $-$16:06:31.8 & 10439     & x              & $0.50\pm0.14$ & x             & Low   & 2,3 & 3 \\
        \noalign{\smallskip}
        ABELL\,2537          & $0.297$ & 23:08:22.08 & $-$02:11:28.4 & 9372      & 0042341201, 0205330501 & $0.82\pm0.11$ & $0.65\pm0.12$ & Low   & 1,4 & 2 \\
        \noalign{\smallskip}
        ABELL\,2552          & $0.300$ & 23:11:33.23 & $+$03:38:02.5 & 11730, 3288 & x              & $0.66\pm0.12$ & x             & Low   &  &2   \\
        \noalign{\smallskip}
        ABELL\,2631          & $0.271$ & 23:37:38.70 & $+$00:16:15.6 & 11728, 3248 & 0042341301     & $1.69\pm0.14$ & $2.77\pm0.11$ & High   & 1,2,3 & 2 \\
        \noalign{\smallskip}
        ABELL\,2645          & $0.251$ & 23:41:16.64 & $-$09:01:23.4 & 11769$^c$ & x              & $1.63\pm0.28$ & x             & High & & 2 \\
        \hline
         \end{tabular}
         }
      
    \label{tab:cluster sample}
\end{table*}

%%%%%%%%%X ray ANALYSIS%%%%%%%%
\section{X-ray Data analysis}
\label{xray_an}

The \emph{Chandra}/ACIS-I and \emph{XMM-Newton}/EPIC observations were
analysed following a similar procedure in which we binned all events  in sky coordinates  and energy, using new analytical
models for the background treatment. The \emph{Chandra} ACIS-S observations,
for which a particle background is not yet analytically modelled, were
analyzed using a ``standard technique'' based on the direct
subtraction of a blank field sky background event-list to the target
event list. The details of these data analysis techniques are
described in the following sections.
 
\subsection{\emph{Chandra} data preparation}

To process the \emph{Chandra} data, we use the {\sc CIAO v4.3}
software, with the {\sc CALDB 4.4.3}.  We start from level 1 event
list and we apply the charge transfer inefficiency (CTI) and time
dependent gain corrections. The CTI is due to soft proton damage to the ACIS chips  during the passages of the satellite through 
the EarthÕs radiation belts and it produces a row dependence in gain, event grade, and energy resolution.  \citep{2000ApJ...534L.139T, 2005SPIE.5898..201G}.
We remove then the bad CCD pixels, taking into account hot pixels, and afterglow
events caused from cosmic rays building up charge on the CCD.
Furthermore,  we assign RA and DEC coordinates to each detected event,
applying ``the aspect solution'' file in which the \emph{Chandra}
dither pattern information is listed.

We then recompute the event grades, analyzing the traces left from
events on detector, to distinguish the photons coming from the
cluster from other particles.  
The grade is a number assigned to every event based on which pixels in its neighboring island are above their threshold value.
The processing examines every pixel in the full CCD image and selects as events regions with bias-subtracted pixel values that both exceed the event threshold and are greater than all of the adjacent pixels (i.e., a local maximum). The surrounding neighboring pixels are then compared to the bias-subtracted split-event threshold; those that are above the threshold establish the pixel pattern. 
For  the observations taken in FAINT  telemetry mode, $3\times3$ islands  are used. For VERY FAINT mode observations, we apply additional
background screening by removing events with significantly positive
pixels at the border of the 5 $\times$ 5 event island. 
On the basis of the pattern, the event is assigned a grade.
With the combination of these values is possible to recognize the object events
path and assign it a good ASCA grades (0, 2, 4, 6), while the other
events, to which is assigned a bad ASCA grades (1, 5, and 7), are
removed.  
Then we apply
the latest gain maps available to compute calibrated photon energies.

To remove the flare episodes, we use the light curve filtering procedure described in
\cite{2003ApJ...583...70M}. We extract the  background light curve in a CCD far from cluster in the
[0.3-12] keV band for Front Illuminated chips, in the [2.5-7] kev band
for S3, and in the [2.5-6] keV band for S1. We bin the light curve with binsize = 259.28 for FI chips and 0.3-12 kev band, or
binsize = 1037.12  for S3 and 2.5-7 kev band, and calculate the average count rate over the standard GTI, using
3$\sigma$ clipping.
Then the bins above
or below that mean by  $ \> 20\%$ are rejected as recommended in the  Markevitch's COOKBOOK\footnote{http://cxc.harvard.edu/contrib/maxim/acisbg/COOKBOOK}.

\subsection{XMM- Newton data preparation}

The \emph{XMM-Newton} data are pre-processed using the \emph{XMM-Newton}-SAS v11.0.
They are then filtered through spatial and temporal wavelet analyses
in order to remove the contribution of point sources and soft proton
flares, following the scheme described in
\cite{2008A&A...479..307B}. 
To detect and remove high solar flares periods, we analyze light
curves with associated high energy events ([10.0-12.0] keV) and softer
events ([1.0-5.0] keV).  This two-step analysis first enables us to
isolate the most prominent flares at high energy, where ICM brightness
is expected to be negligible, and also detect some of the flares with
softer spectra. For each curve, we use a ``B3-spline
\`a-trous'' wavelet algorithm to remove both intervals with no data and  the points 
exceeding a 2$\sigma$ significance the light-curve fluctuation.

\subsection{\emph{Chandra} ACIS-I and \emph{XMM-Newton} EPIC observations}

\emph{Chandra} ACIS-I and \emph{XMM-Newton} EPIC observations are
analyzed following a procedure described in \citet[][hereafter
  BM08]{2008A&A...479..307B} where all event-lists are merged in sky
coordinates and energy building a single energy position
 event cube.  This event cube is first used to identify
point-sources, that are masked during the analysis, adopting an object
separation algorithm derived from the multi-scale vision model
\citep{1995SignalProcessing...46...345}.

Following an approach detailed in BM08, we similarly binned two
quantities useful for imaging and spectroscopy: the effective exposure
time and the estimate of a background noise level.  The effective exposure includes spatial variations of the mirror
effective area and detector quantum efficiency at low energy
resolution, the CCD gaps and bad pixels, and a correction for the
telescope motion. These quantities
are extracted from the \emph{Chandra} Calibration data base (CALDB 4.4.3),
the EPIC Currrent Calibration files (CCFs) and the events list. The
background model array is modeled as the sum of components accounting
for the Galactic foreground and the Cosmic X-ray Background, as well
as false photon detections due to charged particle-induced and
out-of-time events. More details about this multi-component background modeling are provided in the following.

\subsubsection{Galactic foreground and the Cosmic X-ray Background}
\label{Bg_descr}

The Cosmic X-ray background is modeled with an absorbed power law of
index $\gamma$= 1.42 \citep[see e.g.][] {2002A&A...389...93L}, while
the Galactic foregrounds are modeled by the sum of two absorbed
thermal components accounting for the Galactic Ôtransabsorption
emission \citep[TAE, kT1 = 0.099 keV, kT2 = 0.248 keV,
    see][]{2000ApJ...543..195K} and an unabsorbed thermal component
accounting for the local hot bubble \citep[LHB, $kT_{LHB}$ = 0.1 keV, see
e.g.][]{2000ApJ...543..195K}. Being associated with real photon
detections, these components are corrected for the effective areas of
each telescope. In most cases, these components are modeled in an
outer annulus of the field-of-view corresponding to r$ \ge
r_{500}$, where the cluster emissivity itself is negligible. As for
the few nearest clusters in our sample for which the cluster
emissivity is not spatially separable from its background, we
constrain the emissivities of each background component from a joint
fit of the cluster emissivity and temperature. We check the goodness 
of such a multi-component fit, verifying that cluster temperatures thus measured in
an outer annulus of the field-of-view are lower than
the average cluster temperature, as is commonly observed in clusters
allowing a full cluster-foreground spatial separation \citep[see
  e.g.][] {2007A&A...461...71P, 2008A&A...486..359L}.

\subsubsection{Analytic  particle background model for ACIS-I 
very faint mode}
\label{Part_bg}

For \emph{Chandra} ACIS-I VF particle background we use the new
particle background model proposed by \cite{Bart2014}.  Here we describe the main features of the model.

Related to the interaction of highly energetic particle with the detector, the ACIS-I instrumental background is spectrally characterised by the superposition of several
fluorescence emission lines $L(E,y)$ onto a continuum $C(E,y)$:
\be 
F_i(E,y) = C_i(E,y) + L_i(E,y),
\label{eq:generale}
\ee 
where E is the energy, $y$ is the CHIPY coordinate and $i$ is the CCD chip index with $i=[0,1,2,3]$.
To isolate its flux from any sky component, an analytical model of the continuum ([0.3-0.7] keV, [4-5.8] keV and [6-7] keV bands)  has been fitted 
to stowed observations performed in the so called very faint mode, and gathered over the 8 years D+E period starting in 2001.
The  image in the continuum band of the stowed background shows a spatial correlation of the 
flux of this continuum with the chip readout direction \citep{Bart2014}. To characterize the spectrum of the continuum emission, it is been extracted, for each CCD, the spectrum in four adjacent strip regions along $y$ direction: each strip region is a rectangle with ($\Delta$x,$\Delta$y) = (1024, 256) pixels.   These spectra show the same shape but  different normalization. The continuum, therefore, is modeled with a power law plus an exponential (position independent) multiplied by the normalization factor that accounts for the observed gradient $\alpha_i$:
\be C_i(E,y)=\alpha_i(y) \left( B_1 e^{-B_2 E} + B_3 E^{-B_4}
\right),
 \label{eq:continuonuovo}
\ee
 where $B_1=0.1493$, $B_2=3.8106$, $B_3=0.0859$, $B_4=0.0292$ are found fitting the equation \ref{eq:continuonuovo} to the overall spectrum extracted from all the CCD using the continuum band.

For fidelity to the observational configuration as well as statistical purposes, the  emission lines are subsequently
fitted to blank sky observations of the D+E period ($t_{\rm exp}\simeq 1.5\times10^6{\rm sec}$), jointly to all background components.
A total of 11 emission lines are found, 6 of which  being position and energy dependent.
They are modeled as: 
\be
\label{eqn:model_lines}
L_{i}(E,y) = D(E) + \sum^{3}_{\ell=1}S_{\ell,i}(E,y) 
\ee 
to take into
account both the contribution from pure emission lines and the effect
related to the over-correction of Charge Transfer Inefficiency (CTI)
which splits 3 emission lines into a mother plus a daughter line
\citep{2006ApJ...645...95H}. In particular, $L$ is the group of
emission lines which remain constant along all the CCDs while
$S_{\ell,i}(E,y)$ is the group of line which varies spatially along
the chip.  We normalize the background in the [9.5-10.6] keV energy band
as suggested in \citep{Bart2014}. This band is chosen: i) because at this energy the ACIS-I effective area is a factor of $\sim100$ lower than at its peak and ii) to minimize the effect related to prediction error for the mother-daughter line system at E $\>$ 9keV.

The median value of the fraction of the cleaned events in the band [0.5-2.5]keV attributed 
to the total background component (Galactic foreground + Cosmic X-ray Background + particle background)
for our sample is of  $\sim \%20, 50\% and \%80$ at the three radii $r_{2500}$, $r_{1000}$ and $r_{500}$ considered 
in this work (see section \ref{results}).

\subsubsection{Analytic particle background model for
    \emph{XMM-Newton} EPIC}

For   \emph{XMM-Newton} EPIC particle background we use the 
particle background model described in \cite{2013ApJ...764...82B}.

The \emph{XMM-Newton} EPIC particle background is modeled from
observations performed in the Filter Wheel Closed (FWC) position
during revolutions 230 to 2027 as for the EPIC-MOS cameras, and 355 to
1905 as for the EPIC-PN camera. Following an approach proposed in
e.g. \cite{2008A&A...478..575K} or \cite{2008A&A...486..359L}, this
model sums a quiescent continuum to a set of florescence emission
lines convolved with the energy response of each detector. Depending
on the solar flare contamination, it is occasionally completed with
residual emission associated with soft protons. To account for
two different spectral shapes in the soft and hard bands, the
quiescent continuum is modeled as the product of a power law with an
inverted error function increasing in the soft band. We set the
emission line energies to the values reported in
\cite{2008A&A...486..359L}, while the soft proton residual is modeled
using an additional power law.  The EPIC-MOS quiescent continuum
exhibits a small emissivity gradient along the RAWY coordinate, which
has been measured and taken into account in the model. This effect is
presumably due to differences in the collecting areas of the imaging
and readout detector regions. The fluorescence lines exhibit a more
complex spatial variation \citep[see e.g.][]
{2002A&A...389...93L,2008A&A...478..575K}. We modeled the emissivity
distribution of the most prominent lines \footnote{Namely the Al, Si
  and Cu, Ni complexes for the EPIC-MOS and EPIC-PN cameras,
  respectively.} from the wavelet filtering of a set of FWC event
images closely straddling each line energy.

Also for XMM-Newton observations, the median value of the fraction of  the cleaned events in the band [0.5-2.5]keV attributed 
to the total background component is of  $\sim \%20, 50\% and \%80$ at the three radii $r_{2500}$, $r_{1000}$ and $r_{500}$ (see section \ref{results}).

\subsection{\emph{Chandra} ACIS-S observations}

The \emph{Chandra} ACIS-S particle background is not yet analytically
modeled, so we analyze ACIS-S observations by means of the direct
subtraction of a blank field sky background event-list from the target
event list, as described, e.g. in \cite{2005ApJ...628..655V}.  The
blank field sky background is processed identically to the cluster
data and reprojected onto the sky using the aspect information from
the cluster pointing.  We renormalize the blank field sky to the
background in each observation, considering a region of the ACIS field
of view free from cluster emission (mainly ACIS-S1 for ACIS-S
observations) and a spectral band (9.5-12 keV) where the \emph{Chandra}
effective area is nearly zero; therefore, all the observed flux is due
to the particle background.  This is possible because the ACIS
background above 9 keV is dominated by events from the charged
particles, and its spectrum is very stable despite secular and
short-term variations of up to 30\% in its intensity.

In addition to the particle-induced background, we check whether the
diffuse soft X-ray background could be an important factor in our
observations and whether adjustments are required. For each
observation, we follow the procedure of \cite{2005ApJ...628..655V}:
extracting a spectrum in the source-free regions of the detector,
subtracting the renormalized blank-field background, and fitting the
residuals in XSPEC (in the [0.4-2.0] keV band) with an unabsorbed
MEKAL model, whose normalization was allowed to be negative. The
obtained best-fit model is therefore included as an additional
component in the spectral fits (with its normalization scaled by the
area).

\subsection{ACIS-I and \emph{XMM-Newton} EPIC  Imaging and Spectroscopy}
\label{ACIS-I and XMM-Newton EPIC Imaging and Spectroscopy}

For \emph{Chandra} ACIS-I and \emph{XMM-Newton} EPIC data, the imaging and the
spectral analysis is performed as in BM08.  Surface brightness
profiles are computed from soft energy ([0.5--2.5keV]) photon images,
corrected for background and effective area.  The surface  brightness
profile is extracted in concentric annuli centered on the centroid of
the cluster. The maximum radius is set to $r_{500}$, which is
estimated using the r-$Y_x$ scaling relation, if the field of view
permits. The minimum radius is chosen to exclude the cool cores or
internal regions with evidence of isophotal asymmetry, likely due to
merger events.

Spectra are also extracted in annuli around the cluster centroid. The
radial boundaries are chosen so that any annulus contains $\ge 3000$
counts. For each cluster, the number of annuli  of the temperature profile is listed in the Tab.~\ref{tab:cluster sample}. Each spectrum is binned in energy to have at least 20 counts
per bin.   We fit the spectra in the [0.7--10keV]
band to an absorbed single-temperature thermal model (WABS(APEC)),
where the metallicity and the temperature is allowed to be free in
each annulus.  In most cases the absorbing column density $N_H$, is
fixed at the value provided by \cite{1990ARA&A..28..215D} radio
surveys, but we always check that it is consistent with the observed
spectrum. For each extraction region and
detector, this spectral model is added to the expected background
spectrum (see also sect. \ref{Bg_descr}), altered by the effective exposure, multiplied by the mirror effective area and
detector quantum efficiency and convolved by a redistribution
function of the photon energies. This
function has been computed via an averaging of the redistribution
functions relative to each event position and registration date. To
reduce the computation time, these latter functions have been
pre-tabulated in detector coordinates and rebinned within our energy
axes. More precisely, we computed them within 128 tiles of the ACIS-I
CCDs, 3 regions of the EPIC-MOS and 10 regions of the EPIC-PN
detectors.  This was undertaken by using the \emph{Chandra} Interactive
Analysis of Observations (CIAO) software and CALDB 4.4.3, and through
an energy rebinning of the canned redistribution matrixes provided in
the \emph{XMM-Newton} calibration data base.

\subsection{CIAO Imaging and Spectroscopy}

For ACIS-S observations, the imaging and the spectral analysis is
performed using the {\sc CIAO} software.  Starting from the new processed
event file, we build images in the [0.5--2.5keV] energy band, which
maximizes the cluster to background flux ratio in \emph{Chandra} data.  We
subtract the blank-field background from the cluster images and create
flat-fielded images using exposure maps that include corrections for
CCD gaps and vignetting.  Point-like sources are removed with a
detection routine based on the wavelet decomposition technique
documented in \cite{1998ApJ...502..558V}. The point sources are
identified using the small scales of the wavelet decomposition and the
corresponding regions are masked out from all further analysis.

We then extract the  surface brightness and temperature profiles in
region chosen similarly for ACIS-I (that is the camera of reference having a smaller effective area and so, usually, less total counts) and XMM-Newton data as described in \S\ref{ACIS-I and XMM-Newton EPIC Imaging and Spectroscopy}. 
During the spectrum
extraction, the background spectrum is taken from the renormalized
blank-field observations using the same region of the source.  We fit
the [0.7-10] keV spectra to an absorbed single-temperature thermal
model (WABS(MEKAL)), where the metallicity and the temperature are
allowed to be free in each annulus, and the absorbing column
density NH is fixed at the value provided by
\cite{1990ARA&A..28..215D} radio survey.
  
\section{Density, 3-D temperature and mass profiles} 
\label{profiles}

The surface brightness and projected temperature profiles are used to
derive density and 3-D temperature profiles.  This is performed using
the ``forward'' method described in, e.g., \cite{2010A&A...514A..93M}.
In particular, following \cite{2005ApJ...628..655V}, we assume a
modified $\beta$-model profile for the density distribution, to
accommodate a power-law behavior in the center and the observed
steepening of the surface brightness in the outskirts
\citep[e.g][]{1999ApJ...525...47V,2005A&A...439..465N,2009A&A...496..343E},
with the addition of a second $\beta-$model to better reproduce the
core profile:
\be 
{\large  n_{p}n_{e}(r)\equiv\frac{n_{0}^{2}(r/r_{c})^{-\alpha}}{[1+{({\frac{r}{r_{c}}})}^{2}]^{3\beta-\alpha/2}}+\frac{n_{02}^{2}}{[1+{({\frac{r}{r_{c2}}})}^{2}]^{3\beta_{2}}}}\ , 
\label{eq:npne} 
\ee 
where $\alpha$, $\beta$, $\beta_{2}$, $n_{0}$, $n_{02}$, $r_{c}$ and
$r_{c2}$ are free parameters.  These profiles are then projected along
the line of sight, in cylindric shells matching the radial bins of the surface brightness profiles.  For \emph{XMM-Newton} data, this
projection includes convolution with the instrument PSF.  The
temperature profile is modeled with a power-law with the form:
\be 
T_{3D}(r)=T_{0}\frac{(r/r_{t})^{-a}}{[1+(\frac{r}{r_{t}})^{2}]^{c/2}}.
\label{eq:T_3d} 
\ee
The temperature profiles are then projected along the line of sight
using the ``\textit{spectroscopic-like temperature}"
\citep{2004MNRAS.354...10M}:
\be
T_{sl}=\frac{\int WT_{3D}\, dV}{\int W\, dV}
\label{eq:Tsl}
\ee where \be W=\frac{n_pn_e}{T_{3D}^{3/4}}.  \ee The best-fit
parameters are determined from a $\chi^2$ minimization technique by
comparing the projected quantities derived from the models with
the observations.

When the low number of counts do not allow us to have a temperature profile with more than two annuli, we assume a constant profile with a = c = $r_{t}$ = 0 in the formula \ref{eq:T_3d}.

\subsection{Mass derivation}

The density and 3-D temperature profiles are used to estimate the
total gravitating mass assuming the hydrostatic equilibrium condition
\citep{Sar1988Cambridge}:
\be\begin{split}
M(r)=&-\frac{2.22\times10^{13} M_{\odot}}{\mu} T_{3D}(r)r \times\\
&\times\left(\frac{\textup{dln}\rho_{g}(r)}{\textup{dln}r}+\frac{\textup{dln}T_{3D}(r)}{\textup{dln}r}\right)
\label{eq:totmass}
\end{split}
\ee
where $\mu$ is the mean molecular weight in a.m.u.  taken to be 0.5964
and $\rho_{g}(r)$ is the gas density profile. The T3D is in units of KeV and $r$ in units of Mpc. The gas density is
derived from the ICM particle number density profile, given directly
by the analytic fit to the projected emission measure profile. For a
primordial He abundance and abundances of heavier elements $Z=0.3
Z_{\odot}$, $\rho_{g}(r)=1.24\sqrt{n_pn_e(r)}*m_p$, with $m_p$ being
the proton mass.

Given M(r), we can calculate the total matter density profile, $\rho_{tot} =(4\pi r^2)^{-1} dM/dr$. 
The total mass is derived from a complex combination of parameters used
in the models of $\rho_{g}(r)$ and $T_{3D}(r)$, with their
uncertainties estimated with the bootstrap method.  The observed data
are used to generate a set of random realizations, assuming a Poisson
distribution for the total counts in each annulus and a gaussian distribution
for the projected temperature. Each realization is then fitted by the
model described in the previous sections. The best fit and the errors
are the mean value and the standard deviation of the resulting fits.

Our analytic model for  $T_{3D}(r)$ allows very steep gradients. 
In some cases, such profiles are formally consistent with the observed projected temperatures 
because projection produces steep gradients. 
However, large values of $dT_{3D}/dr$  often lead to unphysical mass estimates,
for example, the profiles with $\rho_{tot} < 0$ at some radius. 
We eliminated this problem in the parametric bootstrap simulations by accepting only those realizations
in which the best-fit  $T_{3D}(r)$ leads to $\rho_{tot}$ $>$ $\rho_{gas}$ in the radial range covered by the data and discarding the others.

 Therefore, the strength of the forward method using the analytical functions  \ref{eq:npne} and \ref{eq:T_3d} is to describe a wide range of the possible profiles. In addition, simulations  \citep{2006MNRAS.369.2013R, 2010A&A...514A..93M} show that the resulting mass reconstruction is in agreement with the resulting mass reconstruction obtained by a non parametric-model that uses directly a global deprojection of the data. This indicates that the systematic effect coming from the assumptions of these analytical forms, is negligible for  the mass estimate.

\begin{table*}
  \begin{center}
    \caption[]{Hydrostatic mass measurements}
\resizebox{.9\textwidth}{!} {
    \begin{tabular}{lcccccccccccc}
      \hline
      Cluster & \multicolumn{3}{c}{Radii (Mpc)} & \multicolumn{3}{c}{\emph{Chandra} Masses ($10^{14}M_{\odot}$)} &  \multicolumn{3}{c}{\emph{XMM-Newton} Masses ($10^{14}M_{\odot}$)} \\
      & $r_{2500}$ & $r_{1000}$ & $r_{500}$ & $M_{2500}$ & $M_{1000}$ & $M_{500}$ & $M_{2500}$ & $M_{1000}$ &  $M_{500}$   \\ 
        \noalign{\smallskip}
      \hline
        \noalign{\smallskip}
ABELL\,2697   & 0.51  & 0.84  & 1.20  &    x      &    x      &    x      &     2.43  $\pm$     0.15      &     4.38  $\pm$     0.38      &     6.29  $\pm$   0.65\\
 \noalign{\smallskip}
        ABELL\,0068   & 0.58  & 0.97  & 1.40  &     3.85  $\pm$     1.57      &     7.22  $\pm$     2.79      &    10.77  $\pm$     4.10      &     3.22  $\pm$     0.25      &     6.69  $\pm$     0.53      &    10.44  $\pm$   0.86\\
 \noalign{\smallskip}
        ABELL\,2813   & 0.50  & 0.86  & 1.24  &     2.54  $\pm$     0.60      &     4.99  $\pm$     1.13      &     7.64  $\pm$     1.71      &     2.04  $\pm$     0.25      &     4.07  $\pm$     0.46      &     6.32  $\pm$   0.69\\
 \noalign{\smallskip}
       ABELL\,0115s   & 0.35  & 0.61  & 0.89  &     0.81  $\pm$     0.30      &     1.63  $\pm$     0.47      &     2.54  $\pm$     0.73      &     0.49  $\pm$     0.08      &     1.63  $\pm$     0.14      &     3.26  $\pm$   0.21\\
 \noalign{\smallskip}
        ABELL\,0141   & 0.34  & 0.67  & 1.02  &     0.81  $\pm$     0.51      &     2.33  $\pm$     0.96      &     3.97  $\pm$     1.37      &    x      &    x      &   x    \\
 \noalign{\smallskip}
ZwCl\,0104.4$+$0048   & 0.34  & 0.54  & 0.76  &     0.71  $\pm$     0.03      &     1.16  $\pm$     0.05      &     1.67  $\pm$     0.07      &    x      &    x      &   x    \\
 \noalign{\smallskip}
        ABELL\,0209   & 0.48  & 0.80  & 1.15  &     1.96  $\pm$     0.38      &     3.64  $\pm$     0.70      &     5.47  $\pm$     1.07      &     1.72  $\pm$     0.07      &     3.45  $\pm$     0.12      &     5.45  $\pm$   0.18\\
 \noalign{\smallskip}
        ABELL\,0267   & 0.51  & 0.82  & 1.17  &     2.48  $\pm$     0.80      &     4.15  $\pm$     1.29      &     5.97  $\pm$     1.84      &    x      &    x      &   x    \\
 \noalign{\smallskip}
        ABELL\,0291   & 0.38  & 0.65  & 0.94  &    x      &    x      &    x      &     0.99  $\pm$     0.11      &     1.91  $\pm$     0.33      &     2.92  $\pm$   0.56\\
 \noalign{\smallskip}
        ABELL\,0383   & 0.48  & 0.75  & 1.01  &     2.01  $\pm$     0.48      &     2.95  $\pm$     0.74      &     3.68  $\pm$     0.90      &     1.62  $\pm$     0.09      &     2.48  $\pm$     0.22      &     3.25  $\pm$   0.45\\
 \noalign{\smallskip}
        ABELL\,0521   & 0.27  & 0.77  & 1.23  &     1.70  $\pm$     0.43      &     4.21  $\pm$     0.93      &     6.78  $\pm$     1.48      &     0.34  $\pm$     0.02      &     3.44  $\pm$     0.17      &     7.05  $\pm$   0.34\\
 \noalign{\smallskip}
        ABELL\,0586   & 0.49  & 0.78  & 1.08  &     2.08  $\pm$     0.32      &     3.27  $\pm$     0.63      &     4.42  $\pm$     0.90      &    x      &    x      &   x    \\
 \noalign{\smallskip}
        ABELL\,0611   & 0.51  & 0.84  & 1.20  &     2.68  $\pm$     0.45      &     4.66  $\pm$     0.75      &     6.80  $\pm$     1.08      &    x      &    x      &   x    \\
 \noalign{\smallskip}
        ABELL\,0697   & 0.60  & 1.03  & 1.50  &     4.21  $\pm$     1.07      &     8.53  $\pm$     1.86      &    13.14  $\pm$     2.74      &    x      &    x      &   x    \\
 \noalign{\smallskip}
ZwCl\,0857.9$+$2107   & 0.42  & 0.66  & 0.91  &     1.89  $\pm$     1.49      &     3.65  $\pm$     2.81      &     5.59  $\pm$     3.99      &     1.19  $\pm$     0.11      &     1.81  $\pm$     0.18      &     2.33  $\pm$   0.23\\
 \noalign{\smallskip}
        ABELL\,0750   & 0.47  & 0.72  & 0.96  &     1.81  $\pm$     0.51      &     2.61  $\pm$     0.95      &     3.17  $\pm$     1.22      &    x      &    x      &   x    \\
 \noalign{\smallskip}
        ABELL\,0773   & 0.56  & 0.88  & 1.21  &     3.23  $\pm$     0.55      &     4.92  $\pm$     0.94      &     6.51  $\pm$     1.62      &     3.00  $\pm$     0.27      &     5.48  $\pm$     0.59      &     7.77  $\pm$   0.94\\
 \noalign{\smallskip}
        ABELL\,0781   & 0.31  & 0.73  & 1.13  &     0.77  $\pm$     0.67      &     3.27  $\pm$     1.78      &     6.09  $\pm$     2.87      &     0.61  $\pm$     0.04      &     3.45  $\pm$     0.22      &     6.72  $\pm$   0.45\\
 \noalign{\smallskip}
ZwCl\,0949.6$+$5207   & 0.41  & 0.66  & 0.93  &     1.24  $\pm$     0.08      &     2.03  $\pm$     0.14      &     2.90  $\pm$     0.20      &    x      &    x      &   x    \\
 \noalign{\smallskip}
        ABELL\,0901   & 0.37  & 0.58  & 0.79  &    x      &    x      &    x      &     0.83  $\pm$     0.10      &     1.32  $\pm$     0.22      &     1.68  $\pm$   0.39\\
 \noalign{\smallskip}
        ABELL\,0907   & 0.51  & 0.80  & 1.08  &     2.33  $\pm$     0.38      &     3.50  $\pm$     0.59      &     4.36  $\pm$     0.74      &     2.05  $\pm$     0.12      &     3.65  $\pm$     0.37      &     5.17  $\pm$   0.64\\
 \noalign{\smallskip}
        ABELL\,0963   & 0.55  & 0.85  & 1.14  &     2.68  $\pm$     0.45      &     4.66  $\pm$     0.75      &     6.80  $\pm$     1.08      &     2.41  $\pm$     0.14      &     4.05  $\pm$     0.40      &     5.60  $\pm$   0.71\\
 \noalign{\smallskip}
ZwCl\,1021.0$+$0426   & 0.55  & 0.89  & 1.26  &     3.27  $\pm$     0.41      &     5.44  $\pm$     0.67      &     7.83  $\pm$     0.96      &     2.89  $\pm$     0.06      &     4.76  $\pm$     0.10      &     6.82  $\pm$   0.14\\
 \noalign{\smallskip}
        ABELL\,1423   & 0.54  & 0.85  & 1.18  &     2.99  $\pm$     1.05      &     4.50  $\pm$     1.06      &     6.02  $\pm$     1.53      &    x      &    x      &   x    \\
 \noalign{\smallskip}
        ABELL\,1451   & 0.50  & 0.92  & 1.36  &    x      &    x      &    x      &     2.30  $\pm$     0.66      &     5.51  $\pm$     0.89      &     8.97  $\pm$   2.18\\
 \noalign{\smallskip}
 RXC\,J1212.3$-$1816  & 0.36  & 0.57  & 0.76  &    x      &    x      &    x      &     0.93  $\pm$     0.21      &     1.42  $\pm$     0.26      &     1.67  $\pm$   0.31\\
 \noalign{\smallskip}
 ZwCl\,1231.4$+$100   & 0.21  & 0.76  & 1.23  &     0.19  $\pm$     0.12      &     3.22  $\pm$     0.80      &     6.82  $\pm$     1.29      &    x      &    x      &   x    \\
 \noalign{\smallskip}
        ABELL\,1682   & 0.47  & 0.84  & 1.24  &     2.06  $\pm$     0.98      &     4.64  $\pm$     2.00      &     7.35  $\pm$     3.06      &    x      &    x      &   x    \\
 \noalign{\smallskip}
        ABELL\,1689   & 0.71  & 1.11  & 1.51  &     6.22  $\pm$     0.52      &     9.51  $\pm$     0.93      &    12.09  $\pm$     1.54      &     5.50  $\pm$     0.40      &     9.01  $\pm$     0.96      &    11.98  $\pm$   1.94\\
 \noalign{\smallskip}
        ABELL\,1758   & 0.39  & 0.88  & 1.38  &    x      &    x      &    x      &     1.11  $\pm$     0.19      &     5.23  $\pm$     0.60      &    10.21  $\pm$   1.54\\
 \noalign{\smallskip}
        ABELL\,1763   & 0.50  & 0.89  & 1.33  &     2.38  $\pm$     0.63      &     5.28  $\pm$     1.27      &     8.61  $\pm$     1.96      &     1.92  $\pm$     0.15      &     4.10  $\pm$     0.34      &     6.60  $\pm$   0.56\\
 \noalign{\smallskip}
        ABELL\,1835   & 0.68  & 1.11  & 1.57  &     5.81  $\pm$     0.53      &    10.14  $\pm$     1.56      &    14.52  $\pm$     2.90      &     4.78  $\pm$     0.18      &     9.24  $\pm$     0.60      &    14.04  $\pm$   1.27\\
 \noalign{\smallskip}
        ABELL\,1914   & 0.68  & 1.03  & 1.38  &     5.37  $\pm$     0.92      &     7.48  $\pm$     1.21      &     9.11  $\pm$     1.59      &     4.59  $\pm$     0.59      &     6.59  $\pm$     0.90      &     8.08  $\pm$   1.00\\
 \noalign{\smallskip}
ZwCl\,1454.8$+$2233   & 0.46  & 0.75  & 1.06  &     1.89  $\pm$     0.36      &     3.28  $\pm$     0.91      &     4.61  $\pm$     1.33      &     1.58  $\pm$     0.07      &     2.60  $\pm$     0.22      &     3.65  $\pm$   0.42\\
 \noalign{\smallskip}
        ABELL\,2009   & 0.55  & 0.91  & 1.28  &     2.87  $\pm$     0.58      &     5.08  $\pm$     1.33      &     7.33  $\pm$     2.47      &    x      &    x      &   x    \\
 \noalign{\smallskip}
ZwCl\,1459.4$+$4240   & 0.44  & 0.73  & 1.08  &     1.89  $\pm$     1.49      &     3.65  $\pm$     2.81      &     5.59  $\pm$     3.99      &     1.68  $\pm$     0.10      &     3.54  $\pm$     0.22      &     5.65  $\pm$   0.36\\
 \noalign{\smallskip}
RXC\,J1504.1$-$0248   & 0.61  & 1.01  & 1.47  &     4.16  $\pm$     1.28      &     7.95  $\pm$     3.60      &    13.15  $\pm$     7.11      &     3.65  $\pm$     0.15      &     7.10  $\pm$     0.42      &    10.93  $\pm$   0.82\\
 \noalign{\smallskip}
        ABELL\,2111   & 0.47  & 0.80  & 1.17  &     2.01  $\pm$     0.70      &     3.94  $\pm$     1.27      &     5.99  $\pm$     1.89      &    x      &    x      &   x    \\
 \noalign{\smallskip}
        ABELL\,2204   & 0.71  & 1.10  & 1.49  &     6.14  $\pm$     0.90      &     9.08  $\pm$     1.45      &    11.16  $\pm$     1.75      &     4.90  $\pm$     0.39      &     8.02  $\pm$     0.81      &    10.66  $\pm$   1.72\\
 \noalign{\smallskip}
        ABELL\,2219   & 0.71  & 1.20  & 1.75  &     2.68  $\pm$     0.45      &     4.66  $\pm$     0.75      &     6.80  $\pm$     1.08      &     4.25  $\pm$     0.31      &     8.80  $\pm$     0.97      &    14.35  $\pm$   2.04\\
 \noalign{\smallskip}
 RX\,J1720.1$+$2638   & 0.56  & 0.88  & 1.23  &     3.06  $\pm$     0.38      &     4.71  $\pm$     0.81      &     6.38  $\pm$     1.72      &     2.77  $\pm$     0.10      &     4.79  $\pm$     0.33      &     6.97  $\pm$   0.68\\
 \noalign{\smallskip}
        ABELL\,2261   & 0.58  & 0.90  & 1.22  &     3.66  $\pm$     1.00      &     5.46  $\pm$     1.52      &     6.75  $\pm$     1.89      &    x      &    x      &   x    \\
 \noalign{\smallskip}
RXC\,J2102.1$-$2431   & 0.43  & 0.70  & 1.00  &    x      &    x      &    x      &     1.35  $\pm$     0.13      &     2.36  $\pm$     0.34      &     3.52  $\pm$   0.61\\
 \noalign{\smallskip}
 RX\,J2129.6$+$0005   & 0.48  & 0.76  & 1.08  &     1.98  $\pm$     0.25      &     3.25  $\pm$     0.39      &     4.65  $\pm$     0.56      &     1.81  $\pm$     0.07      &     2.95  $\pm$     0.11      &     4.22  $\pm$   0.16\\
 \noalign{\smallskip}
        ABELL\,2390   & 0.64  & 1.10  & 1.59  &     2.68  $\pm$     0.45      &     4.66  $\pm$     0.75      &     6.80  $\pm$     1.08      &     4.10  $\pm$     0.30      &     8.80  $\pm$     0.92      &    13.67  $\pm$   2.09\\
 \noalign{\smallskip}
        ABELL\,2485   & 0.46  & 0.78  & 1.11  &     1.80  $\pm$     0.43      &     3.58  $\pm$     1.16      &     5.32  $\pm$     2.08      &    x      &    x      &   x    \\
 \noalign{\smallskip}
        ABELL\,2537   & 0.51  & 0.83  & 1.18  &     2.73  $\pm$     0.74      &     4.60  $\pm$     1.21      &     6.65  $\pm$     1.73      &     2.51  $\pm$     0.23      &     4.73  $\pm$     0.45      &     7.20  $\pm$   0.73\\
 \noalign{\smallskip}
        ABELL\,2552   & 0.53  & 0.88  & 1.25  &     3.04  $\pm$     0.70      &     5.34  $\pm$     1.14      &     7.81  $\pm$     1.64      &    x      &    x      &   x    \\
 \noalign{\smallskip}
        ABELL\,2631   & 0.44  & 0.82  & 1.20  &     1.70  $\pm$     0.43      &     4.21  $\pm$     0.93      &     6.78  $\pm$     1.48      &     1.96  $\pm$     0.25      &     5.12  $\pm$     0.60      &     8.51  $\pm$   0.98\\
 \noalign{\smallskip}
        ABELL\,2645   & 0.47  & 0.80  & 1.15  &     2.23  $\pm$     1.08      &     4.05  $\pm$     1.80      &     5.98  $\pm$     2.59      &    x      &    x      &   x    \\
 \noalign{\smallskip}
 \noalign{\smallskip}
      \hline
    \end{tabular}
 }
  \label{tab:massex}
\end{center}
\end{table*}

\begin{table*}
  \begin{center}
    \caption[]{Gas mass measurements}
     \resizebox{.8\textwidth}{!} {
    \begin{tabular}{lcccccccccccc}
      \hline
        \noalign{\smallskip}
      Cluster      &  \multicolumn{3}{c}{\emph{Chandra} Gas Masses ($10^{14}M_{\odot}$)} &  \multicolumn{3}{c}{\emph{XMM-Newton} Gas Masses($10^{14}M_{\odot}$)} \\
      & $M_{\rm gas,2500}$ & $M_{\rm gas,1000}$& $M_{\rm gas,500}$ & $M_{\rm gas,2500}$ & $M_{\rm gas,1000}$&  $M_{\rm gas,500}$   \\ 
        \noalign{\smallskip}
      \hline
        \noalign{\smallskip}
          ABELL\,2697   &                       x   &                       x   &                       x   &       0.275 $\pm$ 0.009   &       0.567 $\pm$ 0.022   &                    0.880 $\pm$ 0.037 \\
 \noalign{\smallskip}
        ABELL\,0068   &       0.322 $\pm$ 0.078   &       0.619 $\pm$ 0.111   &       0.903 $\pm$ 0.135   &       0.316 $\pm$ 0.001   &       0.633 $\pm$ 0.004   &                    0.928 $\pm$ 0.008 \\
 \noalign{\smallskip}
        ABELL\,2813   &       0.300 $\pm$ 0.041   &       0.642 $\pm$ 0.069   &       1.014 $\pm$ 0.092   &       0.290 $\pm$ 0.005   &       0.625 $\pm$ 0.006   &                    0.999 $\pm$ 0.007 \\
 \noalign{\smallskip}
       ABELL\,0115s   &       0.091 $\pm$ 0.024   &       0.270 $\pm$ 0.049   &       0.546 $\pm$ 0.089   &       0.086 $\pm$ 0.000   &       0.264 $\pm$ 0.001   &                    0.519 $\pm$ 0.003 \\
 \noalign{\smallskip}
        ABELL\,0141   &       0.083 $\pm$ 0.036   &       0.295 $\pm$ 0.064   &       0.550 $\pm$ 0.084   &                       x   &                       x   &                                     x \\
 \noalign{\smallskip}
ZwCl\,0104.4$+$0048   &       0.236 $\pm$ 0.004   &       0.161 $\pm$ 0.003   &       0.236 $\pm$ 0.004   &                       x   &                       x   &                                     x \\
 \noalign{\smallskip}
        ABELL\,0209   &       0.244 $\pm$ 0.028   &       0.564 $\pm$ 0.059   &       0.972 $\pm$ 0.094   &       0.226 $\pm$ 0.001   &       0.531 $\pm$ 0.002   &                    0.909 $\pm$ 0.003 \\
 \noalign{\smallskip}
        ABELL\,0267   &       0.232 $\pm$ 0.038   &       0.449 $\pm$ 0.064   &       0.703 $\pm$ 0.094   &                       x   &                       x   &                                     x \\
 \noalign{\smallskip}
        ABELL\,0291   &                       x   &                       x   &                       x   &       0.123 $\pm$ 0.006   &       0.245 $\pm$ 0.018   &                    0.391 $\pm$ 0.031 \\
 \noalign{\smallskip}
        ABELL\,0383   &       0.180 $\pm$ 0.018   &       0.302 $\pm$ 0.026   &       0.425 $\pm$ 0.036   &       0.175 $\pm$ 0.001   &       0.300 $\pm$ 0.001   &                    0.433 $\pm$ 0.002 \\
 \noalign{\smallskip}
        ABELL\,0521   &       0.230 $\pm$ 0.039   &       0.626 $\pm$ 0.068   &       1.027 $\pm$ 0.088   &       0.053 $\pm$ 0.000   &       0.579 $\pm$ 0.001   &                    1.198 $\pm$ 0.003 \\
 \noalign{\smallskip}
        ABELL\,0586   &       0.221 $\pm$ 0.016   &       0.402 $\pm$ 0.033   &       0.600 $\pm$ 0.049   &                       x   &                       x   &                                     x \\
 \noalign{\smallskip}
        ABELL\,0611   &       0.612 $\pm$ 0.039   &       0.406 $\pm$ 0.028   &       0.612 $\pm$ 0.039   &                       x   &                       x   &                                     x \\
 \noalign{\smallskip}
        ABELL\,0697   &       0.423 $\pm$ 0.059   &       0.936 $\pm$ 0.095   &       1.485 $\pm$ 0.127   &                       x   &                       x   &                                     x \\
 \noalign{\smallskip}
ZwCl\,0857.9$+$2107   &       0.190 $\pm$ 0.076   &       0.412 $\pm$ 0.151   &       0.675 $\pm$ 0.187   &       0.154 $\pm$ 0.001   &       0.263 $\pm$ 0.002   &                    0.382 $\pm$ 0.004 \\
 \noalign{\smallskip}
        ABELL\,0750   &       0.154 $\pm$ 0.020   &       0.277 $\pm$ 0.041   &       0.406 $\pm$ 0.061   &                       x   &                       x   &                                     x \\
 \noalign{\smallskip}
        ABELL\,0773   &       0.317 $\pm$ 0.030   &       0.599 $\pm$ 0.056   &       0.907 $\pm$ 0.102   &       0.307 $\pm$ 0.001   &       0.589 $\pm$ 0.002   &                    0.869 $\pm$ 0.003 \\
 \noalign{\smallskip}
        ABELL\,0781   &       0.084 $\pm$ 0.060   &       0.420 $\pm$ 0.121   &       0.783 $\pm$ 0.142   &       0.076 $\pm$ 0.001   &       0.422 $\pm$ 0.001   &                    0.771 $\pm$ 0.004 \\
 \noalign{\smallskip}
ZwCl\,0949.6$+$5207   &       0.315 $\pm$ 0.010   &       0.200 $\pm$ 0.007   &       0.315 $\pm$ 0.010   &                       x   &                       x   &                                     x \\
 \noalign{\smallskip}
        ABELL\,0901   &                       x   &                       x   &                       x   &       0.074 $\pm$ 0.004   &       0.137 $\pm$ 0.010   &                    0.208 $\pm$ 0.020 \\
 \noalign{\smallskip}
        ABELL\,0907   &       0.235 $\pm$ 0.018   &       0.425 $\pm$ 0.030   &       0.623 $\pm$ 0.042   &       0.246 $\pm$ 0.001   &       0.443 $\pm$ 0.002   &                    0.648 $\pm$ 0.004 \\
 \noalign{\smallskip}
        ABELL\,0963   &       0.612 $\pm$ 0.039   &       0.406 $\pm$ 0.028   &       0.612 $\pm$ 0.039   &       0.279 $\pm$ 0.001   &       0.500 $\pm$ 0.003   &                    0.714 $\pm$ 0.003 \\
 \noalign{\smallskip}
ZwCl\,1021.0$+$0426   &       0.422 $\pm$ 0.024   &       0.739 $\pm$ 0.036   &       1.082 $\pm$ 0.048   &       0.473 $\pm$ 0.001   &       0.800 $\pm$ 0.001   &                    1.151 $\pm$ 0.002 \\
 \noalign{\smallskip}
        ABELL\,1423   &       0.216 $\pm$ 0.043   &       0.435 $\pm$ 0.056   &       0.711 $\pm$ 0.095   &                       x   &                       x   &                                     x \\
 \noalign{\smallskip}
        ABELL\,1451   &                       x   &                       x   &                       x   &       0.235 $\pm$ 0.044   &       0.627 $\pm$ 0.051   &                    1.046 $\pm$ 0.102 \\
 \noalign{\smallskip}
 RXC\,J1212.3$-$1816  &                       x   &                       x   &                       x   &       0.069 $\pm$ 0.009   &       0.138 $\pm$ 0.010   &                    0.196 $\pm$ 0.012 \\
 \noalign{\smallskip}
 ZwCl\,1231.4$+$100   &       0.018 $\pm$ 0.011   &       0.374 $\pm$ 0.060   &       0.828 $\pm$ 0.078   &                       x   &                       x   &                                     x \\
 \noalign{\smallskip}
        ABELL\,1682   &       0.180 $\pm$ 0.057   &       0.460 $\pm$ 0.103   &       0.764 $\pm$ 0.137   &                       x   &                       x   &                                     x \\
 \noalign{\smallskip}
        ABELL\,1689   &       0.570 $\pm$ 0.021   &       0.944 $\pm$ 0.036   &       1.285 $\pm$ 0.059   &       0.573 $\pm$ 0.002   &       0.946 $\pm$ 0.004   &                    1.250 $\pm$ 0.007 \\
 \noalign{\smallskip}
        ABELL\,1758   &                       x   &                       x   &                       x   &       0.154 $\pm$ 0.021   &       0.716 $\pm$ 0.041   &                    1.218 $\pm$ 0.062 \\
 \noalign{\smallskip}
        ABELL\,1763   &       0.274 $\pm$ 0.046   &       0.705 $\pm$ 0.093   &       1.229 $\pm$ 0.135   &       0.265 $\pm$ 0.001   &       0.688 $\pm$ 0.003   &                    1.233 $\pm$ 0.005 \\
 \noalign{\smallskip}
        ABELL\,1835   &       0.606 $\pm$ 0.025   &       1.076 $\pm$ 0.065   &       1.554 $\pm$ 0.120   &       0.643 $\pm$ 0.002   &       1.138 $\pm$ 0.003   &                    1.615 $\pm$ 0.007 \\
 \noalign{\smallskip}
        ABELL\,1914   &       0.521 $\pm$ 0.038   &       0.844 $\pm$ 0.054   &       1.163 $\pm$ 0.073   &       0.503 $\pm$ 0.001   &       0.823 $\pm$ 0.003   &                    1.141 $\pm$ 0.004 \\
 \noalign{\smallskip}
ZwCl\,1454.8$+$2233   &       0.222 $\pm$ 0.017   &       0.392 $\pm$ 0.040   &       0.578 $\pm$ 0.060   &       0.223 $\pm$ 0.000   &       0.403 $\pm$ 0.001   &                    0.608 $\pm$ 0.002 \\
 \noalign{\smallskip}
        ABELL\,2009   &       0.261 $\pm$ 0.024   &       0.481 $\pm$ 0.049   &       0.708 $\pm$ 0.082   &                       x   &                       x   &                                     x \\
 \noalign{\smallskip}
ZwCl\,1459.4$+$4240   &       0.190 $\pm$ 0.076   &       0.412 $\pm$ 0.151   &       0.675 $\pm$ 0.187   &       0.180 $\pm$ 0.001   &       0.415 $\pm$ 0.002   &                    0.684 $\pm$ 0.003 \\
 \noalign{\smallskip}
RXC\,J1504.1$-$0248   &       0.485 $\pm$ 0.066   &       0.858 $\pm$ 0.179   &       1.301 $\pm$ 0.337   &       0.525 $\pm$ 0.001   &       0.928 $\pm$ 0.002   &                    1.347 $\pm$ 0.004 \\
 \noalign{\smallskip}
        ABELL\,2111   &       0.181 $\pm$ 0.040   &       0.427 $\pm$ 0.074   &       0.719 $\pm$ 0.110   &                       x   &                       x   &                                     x \\
 \noalign{\smallskip}
        ABELL\,2204   &       0.529 $\pm$ 0.036   &       0.905 $\pm$ 0.062   &       1.279 $\pm$ 0.082   &       0.588 $\pm$ 0.003   &       1.004 $\pm$ 0.002   &                    1.385 $\pm$ 0.008 \\
 \noalign{\smallskip}
        ABELL\,2219   &       0.612 $\pm$ 0.039   &       0.406 $\pm$ 0.028   &       0.612 $\pm$ 0.039   &       0.690 $\pm$ 0.003   &       1.448 $\pm$ 0.010   &                    2.324 $\pm$ 0.023 \\
 \noalign{\smallskip}
 RX\,J1720.1$+$2638   &       0.296 $\pm$ 0.017   &       0.521 $\pm$ 0.038   &       0.771 $\pm$ 0.083   &       0.306 $\pm$ 0.001   &       0.550 $\pm$ 0.001   &                    0.821 $\pm$ 0.003 \\
 \noalign{\smallskip}
        ABELL\,2261   &       0.383 $\pm$ 0.050   &       0.688 $\pm$ 0.088   &       1.004 $\pm$ 0.127   &                       x   &                       x   &                                     x \\
 \noalign{\smallskip}
RXC\,J2102.1$-$2431   &                       x   &                       x   &                       x   &       0.146 $\pm$ 0.006   &       0.280 $\pm$ 0.017   &                    0.450 $\pm$ 0.033 \\
 \noalign{\smallskip}
 RX\,J2129.6$+$0005   &       0.249 $\pm$ 0.015   &       0.475 $\pm$ 0.025   &       0.749 $\pm$ 0.037   &       0.254 $\pm$ 0.000   &       0.476 $\pm$ 0.001   &                    0.744 $\pm$ 0.002 \\
 \noalign{\smallskip}
        ABELL\,2390   &       0.612 $\pm$ 0.039   &       0.406 $\pm$ 0.028   &       0.612 $\pm$ 0.039   &       0.583 $\pm$ 0.003   &       1.225 $\pm$ 0.007   &                    1.861 $\pm$ 0.023 \\
 \noalign{\smallskip}
        ABELL\,2485   &       0.163 $\pm$ 0.020   &       0.354 $\pm$ 0.051   &       0.558 $\pm$ 0.087   &                       x   &                       x   &                                     x \\
 \noalign{\smallskip}
        ABELL\,2537   &       0.242 $\pm$ 0.033   &       0.470 $\pm$ 0.056   &       0.739 $\pm$ 0.081   &       0.242 $\pm$ 0.002   &       0.475 $\pm$ 0.004   &                    0.725 $\pm$ 0.009 \\
 \noalign{\smallskip}
        ABELL\,2552   &       0.316 $\pm$ 0.037   &       0.644 $\pm$ 0.064   &       1.022 $\pm$ 0.094   &                       x   &                       x   &                                     x \\
 \noalign{\smallskip}
        ABELL\,2631   &       0.230 $\pm$ 0.039   &       0.626 $\pm$ 0.068   &       1.027 $\pm$ 0.088   &       0.223 $\pm$ 0.002   &       0.601 $\pm$ 0.005   &                    0.970 $\pm$ 0.008 \\
 \noalign{\smallskip}
        ABELL\,2645   &       0.541 $\pm$ 0.117   &       0.328 $\pm$ 0.079   &       0.541 $\pm$ 0.117   &                       x   &                       x   &                                     x \\
 \noalign{\smallskip}
             \noalign{\smallskip}
      \hline
    \end{tabular}}
    \label{tab:massegasx}
  \end{center}
\end{table*}

\section{X-ray morphology}
\label{sec:morph}

We adopt the X-ray centroid shift as a measure of X-ray morphology
\citep[following:][]{1982ARA&A..20..547F, 1995ApJ...452..522B,
  2006MNRAS.373..881P, 2008ApJS..174..117M, 2010A&A...514A..32B,
  2012ApJ...754..119M}.  The centroid shift parameter \ww\ is  defined
as the standard deviation of the projected separation between the
X-ray peak and the centroid of emission calculated in circular
apertures centered on the X-ray peak. We measure \ww\ (in unit of pixels) for each
cluster in our sample following the method described in
\cite{1995ApJ...447....8M} as implemented by
\cite{2008ApJS..174..117M}.  The analysis is performed on
exposure-corrected and source-masked images in the energy band
[0.7--2.0keV], binned to $1\,\rm{arcsec}^2$ pixels for both
\emph{Chandra} and \emph{XMM-Newton}.  The circular apertures have a
minimum radius equal to $30 \rm{kpc}$, so that the core is excised
from the calculation of the centroid, and the maximum radius decreases
from $r_{500}$ to $0.05 r_{500}$ in steps of 5\% of $r_{500}$.  The
core is included only for the determination of the X-ray peak
position. The error on \ww\ is derived from the standard deviation
observed in cluster images resimulated with Poisson noise, following
\cite{2010A&A...514A..32B}. The \ww\  is then normalized in unit of $10^{-2}r_{500}$.

Previous observational studies of the \ww\ distribution and the
comparison with simulations  \citep[e.g.][]
{2012ApJ...754..119M,2008ApJS..174..117M, 2009A&A...498..361P,
  2010A&A...514A..32B} have adopted a dividing line between ``High-w''
and ``Low-w'' sub-samples at $\wwm=10^{-2}r_{500}$.  ``High-w''
clusters typically show an irregular X-ray morphology due to
cluster-cluster merger activity, while ``Low-w''clusters are
characterized by a regular and more circular X-ray morphology that is
generally associated with clusters being in dynamical equilibrium.
However note that the interpretation of Low-w clusters may be
degenerate to spherical, dynamically relaxed clusters and clusters
that are merging along an axis close to the observer's line of sight
\citep{2012ApJ...754..119M}.  Nevertheless, we adopt this definition
of Low- and High-w sub-samples to aide interpretation of our results,
and comparison of them with the literature. We classify 21/50
clusters as ``High-w'' and 29/50 as ``Low-w'' (Table \ref{tab:cluster
  sample}).  Furthermore, we notice that  the clusters having both  \emph{Chandra} and
\emph{XMM-Newton} data, have \ww values  consistent with each other.

%%%%%%%RESULTS%%%%%%%%%%%%%
\section{Results}
\label{results}
 
We extract electron density, gas mass, hydrostatic mass, and
temperature profiles for all 50 clusters from the X-ray data, using
the techniques from \S\S3~\&~4.  For each cluster, the gas masses and
hydrostatic masses are derived within three different radii,
$r_{\Delta}$, with $\Delta=2500,\,1000$ and 500, where $r_{\Delta}$ is
the clustercentric radius containing the mass
$M_{\Delta}=\Delta\bar{\rho}\frac{4}{3}\pi r_{\Delta}^3$, with
$\bar{\rho}$ the critical density of the Universe at the redshift of
the cluster (Tables~\ref{tab:massex}~\&~\ref{tab:massegasx}).

We describe our main results below, beginning with a comparison of
\emph{Chandra} and \emph{XMM-Newton} measurements of the 21 clusters
that have been observed with both \emph{Chandra}/ACIS-I and
\emph{XMM-Newton} in \S\ref{gas_mass_cxo_xmm}, moving on to a
comparison of our new X-ray measurements and published weak-lensing
mass measurements in \S\ref{xray_wl_mass}, and closing with a
comparison of the X-ray measurements and published measurements of the
Sunyaev-Zeldovich Effect in \S\ref{xray_sz}.

\subsection{Cross-calibration of \emph{Chandra} and
    \emph{XMM-Newton}}\label{gas_mass_cxo_xmm}

\begin{figure*} 
  \begin{center} 
    \leavevmode
    \includegraphics[width=\textwidth]{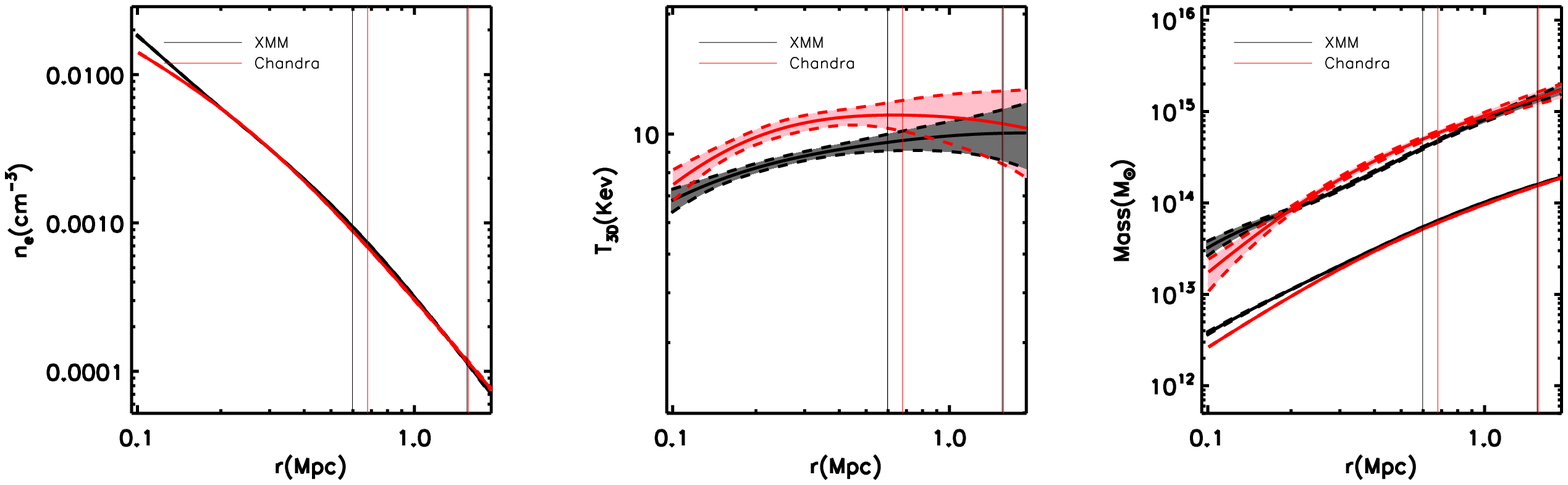}\\
    \includegraphics[width=\textwidth]{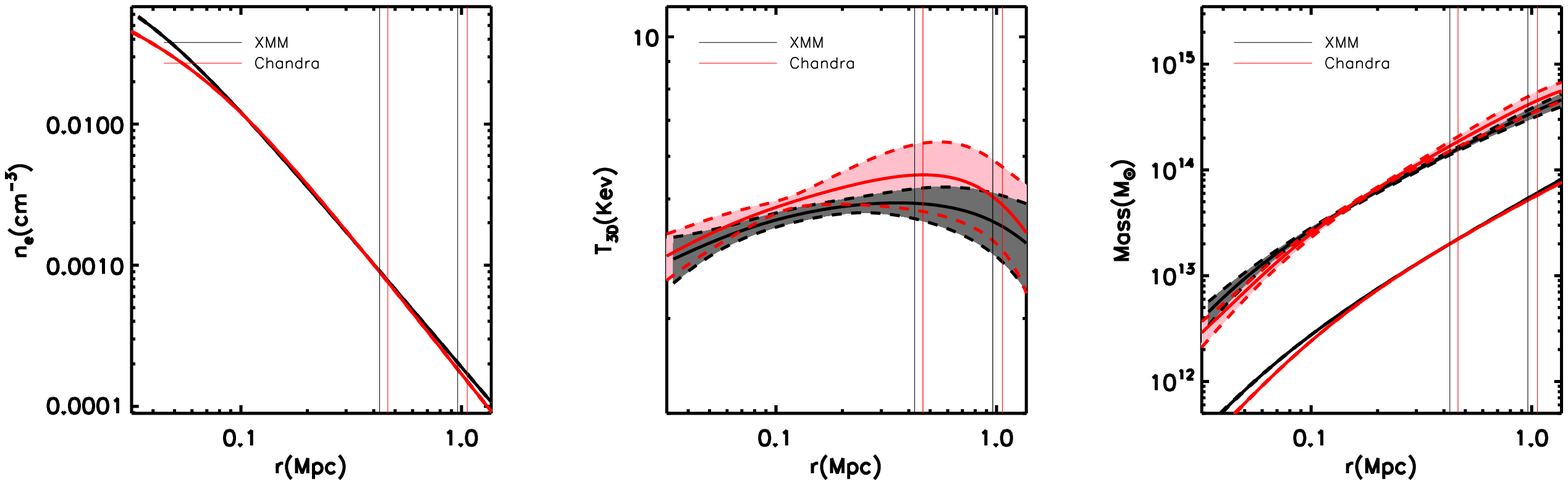}\\
    \includegraphics[width=\textwidth]{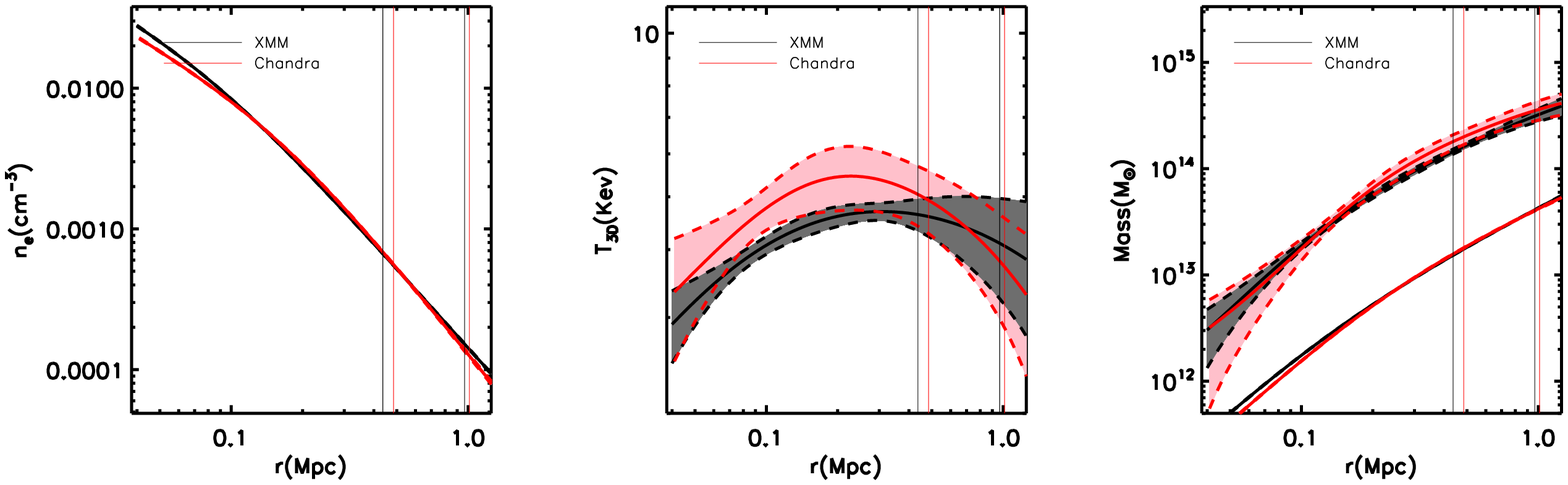}\\
    \caption{Gas density (left), de-projected temperature (centre) and
      gas mass and total hydrostatic mass profiles (right), for three
      respresentative clusters from the High-$\Lx$ sample: ABELL01835
      (top row), ZwCl\,1454.8$+$2233(middle row), and ABELL0383 (bottom
      row).  The vertical lines indicate $r_{2500}$ and $r_{500}$,
      respectively.  \emph{Chandra} is plotted in red; \emph{XMM-Newton} in
      black. The shaded regions and dashed curves delineate the
      respective 1$\sigma$ uncertainties.}
    \label{fig:cxo_xmm_profiles}
  \end{center}
\end{figure*}

The gas density and de-projected temperature profiles from the
\emph{Chandra}/ACIS-I and \emph{XMM-Newton} data agree within the
statistical errors (Figure~\ref{fig:cxo_xmm_profiles}).  The agreement
is particularly good on the scales over which cluster mass is
typically measured for cosmological studies, i.e.\ between $r_{2500}$
and $r_{500}$ -- see vertical lines in
Figure~\ref{fig:cxo_xmm_profiles}.  However, the agreement may be an
artefact of the large uncertainties on these scales.  Moreover,
despite the good agreement within the errors, the \emph{Chandra}-based
temperature profiles generally lie above the \emph{XMM-Newton}-based
profiles, e.g.\ ABELL\,0383 (bottom row of
Figure~\ref{fig:cxo_xmm_profiles}).  We therefore compare the
measurements of the de-projected temperature profiles between the two
satellites and between the different detectors/cameras on-board
\emph{XMM-Newton}.  The details of this rather technical exercise are
presented in the Appendix; we summarize the main points in the
following paragraph.

We extract 4 spectra from the same circular region within $r_{500}$
for ACIS-I, MOS1, MOS2 and PN, excluding flux from point sources, and
gaps between the detectors used in all four instruments by masking
identical regions from all datasets.  We consider the \emph{Chandra}
spectrum as the reference spectrum and fit it in the [0.3-10.0]\,keV
energy band with an absorbed APEC model with temperature, metallicity
and $N_H$ as free parameters. Then, we compare the resulting best fit
model with the spectra obtained from the other cameras.  The average
ratio of temperature measurements between the satellites is measured
to be: $\langle T_{\rm XMM}/T_{\rm
    Chandra}\rangle=0.91\pm0.03$, where $T_{\rm XMM}$ is based on all
three cameras, PN, MOS1, and MOS2.  However this disagreement between
satellites masks the differences between the three \emph{XMM-Newton}
cameras, with MOS1 reporting temperatures $4\%$ higher than the
\emph{XMM-Newton} mean and $8\%$ higher than PN
(Table~\ref{tab:Tratio_cxo_xmm}).  We conclude that, based on their
current calibration files, X-ray temperature measurements currently
suffer $\sim4-9\%$ systematic uncertainty between satellites and
between cameras.  The temperature differences between satellites and
instruments seems to depend on cluster temperature, based on fitting a
two-parameter model to the respective measurements, $T_1$ and $T_2$:
$T_1=A\,T_2^\alpha$.  We obtain best fits of $\alpha=0.81\pm0.17$,
$0.92\pm0.17$, $0.83\pm0.14$, $0.84\pm0.12$ and  $A=1.24\pm0.31$,
$2.50\pm0.42$, $1.29\pm0.39$, $1.24\pm0.31$
 for the PN/ACIS-I, MOS1/ACIS-I,
MOS2/ACIS-I and ALL 3 XMM CAMERAS/ACIS-I combinations respectively.  In each case a larger
  discrepancy is found at higher temperatures.

\begin{figure*}
  \begin{center} \leavevmode
   \includegraphics[width=1.\textwidth]{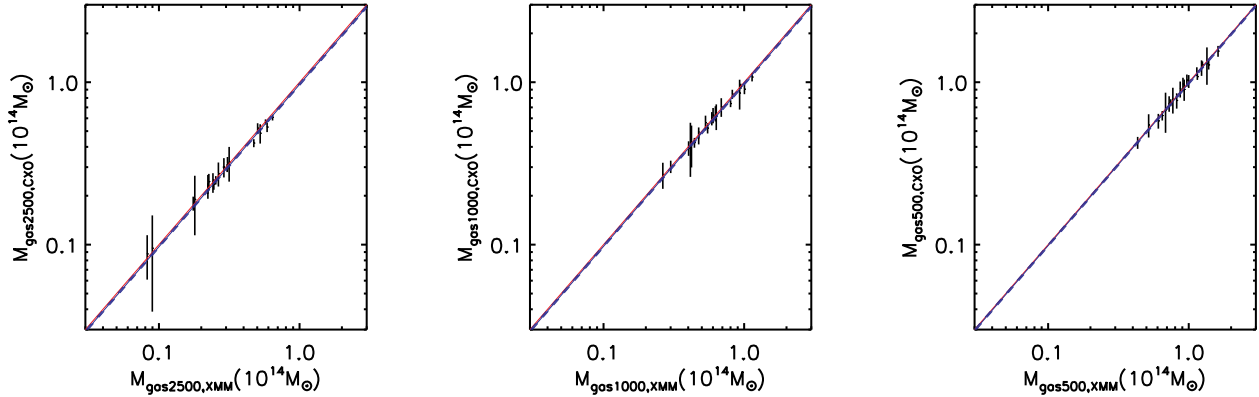}
    \caption{\textit{Left panel:} \emph{Chandra} gas mass versus \emph{XMM-Newton} gas mass
       at $r_{2500}$: the solid blue line is the best fit
      relation with fixed slope = 1; the dashed blue lines are the errors of the best fit
      at $1\sigma$, the solid red line is the expected line
      1:1. \textit{Middle panel:} \emph{Chandra} gas mass versus  \emph{XMM-Newton} gas mass
       at $r_{1000}$, similarly for \emph{Chandra} versus
      \emph{XMM-Newton} gas mass  at $r_{2500}$.. \textit{Right panel:} \emph{Chandra} gas masses versus
      \emph{XMM-Newton} gas mass  at $r_{500}$, with lines similar for $r_{2500}$ and $r_{1000}$.}
    \label{fig:cxo_xmm_ratio_mgas}
  \end{center}
\end{figure*}

\emph{Chandra} and \emph{XMM-Newton} gas mass measurements agree with
each other within the uncertainties
(Figure~\ref{fig:cxo_xmm_ratio_mgas}~\& ~Table~\ref{tab:gas_mass_comp}), based
on fitting the model to the data: $M_{\rm
  CXO}/M_0=a_{\Delta}\,(M_{\rm XMM}/M_0)^{\alpha}$, and minimizing a modified
error-weighted $\chi^2$ statistic \citep{2008MNRAS.384.1567M,
  1992nrfa.book.....P}.  
  In Table~\ref{tab:gas_mass_comp} we report the measured parameter values with two different treatments of the slope ($\alpha$), both as a free parameter of the fit, and with $\alpha$ fixed to 1.
  This agreement holds well at all three
over-density considered: $\Delta=500, 1000, 2500$.  We therefore
conclude that  the satellites
measure the same X-ray flux within the [0.5--2.5keV] energy band to $2\%$ precision.

\begin{figure*}%[htbp]
  \begin{center} \leavevmode
    \includegraphics[width=1.\textwidth]{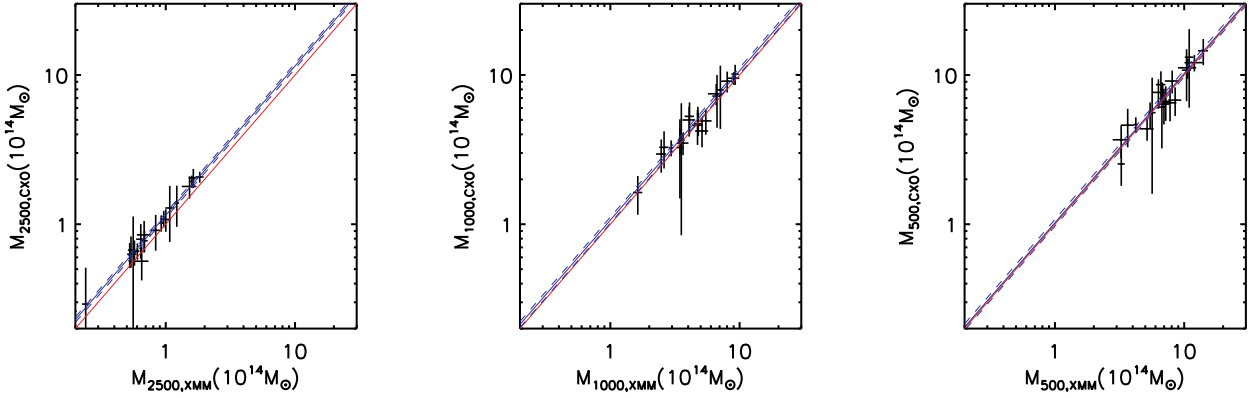}
    \caption{ \textit{Left panel:} \emph{Chandra} mass versus \emph{XMM-Newton}
      mass at $r_{2500}$: the solid blue line is the best fit
      relation with fixed slope = 1; the dashed blue lines are the errors of the best fit
      at $1\sigma$, the solid red line is the expected line
      1:1. \textit{Middle panel:} \emph{Chandra} mass versus \emph{XMM-Newton}
      mass  at $r_{1000}$, similarly for \emph{Chandra} versus
      \emph{XMM-Newton} mass  at $r_{2500}$. \textit{Right
        panel:} \emph{Chandra} mass versus  \emph{XMM-Newton} mass at
      $r_{500}$, with lines similar for $r_{2500}$ and $r_{1000}$.}
    \label{fig:cxo_xmm_ratio}
  \end{center}
\end{figure*}

We find good agreement between the total hydrostatic mass
measurements, albeit with slightly larger uncertainties
(Figure~\ref{fig:cxo_xmm_ratio} \& Table~\ref{tab:masscomp}).  The
agreement is particularly good at $r_{500}$, with
$a_{500}=1.02\pm0.05$ for the fixed-slope model, and deteriorates
slightly towards higher over-densities.  Despite the uncertainties on
the global temperature, the agreement of the the total mass is due to
the agreement of the temperature profiles at larger radii. In fact,
many clusters show a discrepancy of the \emph{XMM-Newton} and
\emph{Chandra} temperature profiles in the inner part, while they are
perfectly in agreement within the errors at larger radii (see
Figure~\ref{fig:cxo_xmm_profiles}).  The apparent $2\sigma$ disagreement
at $r_{2500}$ may be caused by the smearing effect of the
\emph{XMM-Newton} PSF on the temperature profile.  This could
modify the temperature profile of clusters with strong temperature
gradients, for example cool core clusters.  Note that we do not take
account of the PSF when de-projecting the \emph{XMM-Newton}
temperature profile.  We therefore divide the clusters into cool core
and non-cool core sub-samples.  Clusters are classed as cool-core if
the temperature within $r<0.15r_{500}$ is cooler than temperature
within $(0.15r_{500}<r<0.3r_{500}$ at $>2\sigma$ significance.  We
obtain mean mass ratios at $r_{2500}$ of $a_{2500}=1.15\pm0.05$ and
$1.13\pm0.07$ respectively.  This suggests that the PSF is not the
main cause of the discrepancy at $\Delta=2500$.

 \begin{figure*}%[htbp]
  \begin{center} \leavevmode
    \includegraphics[width=1.\textwidth]{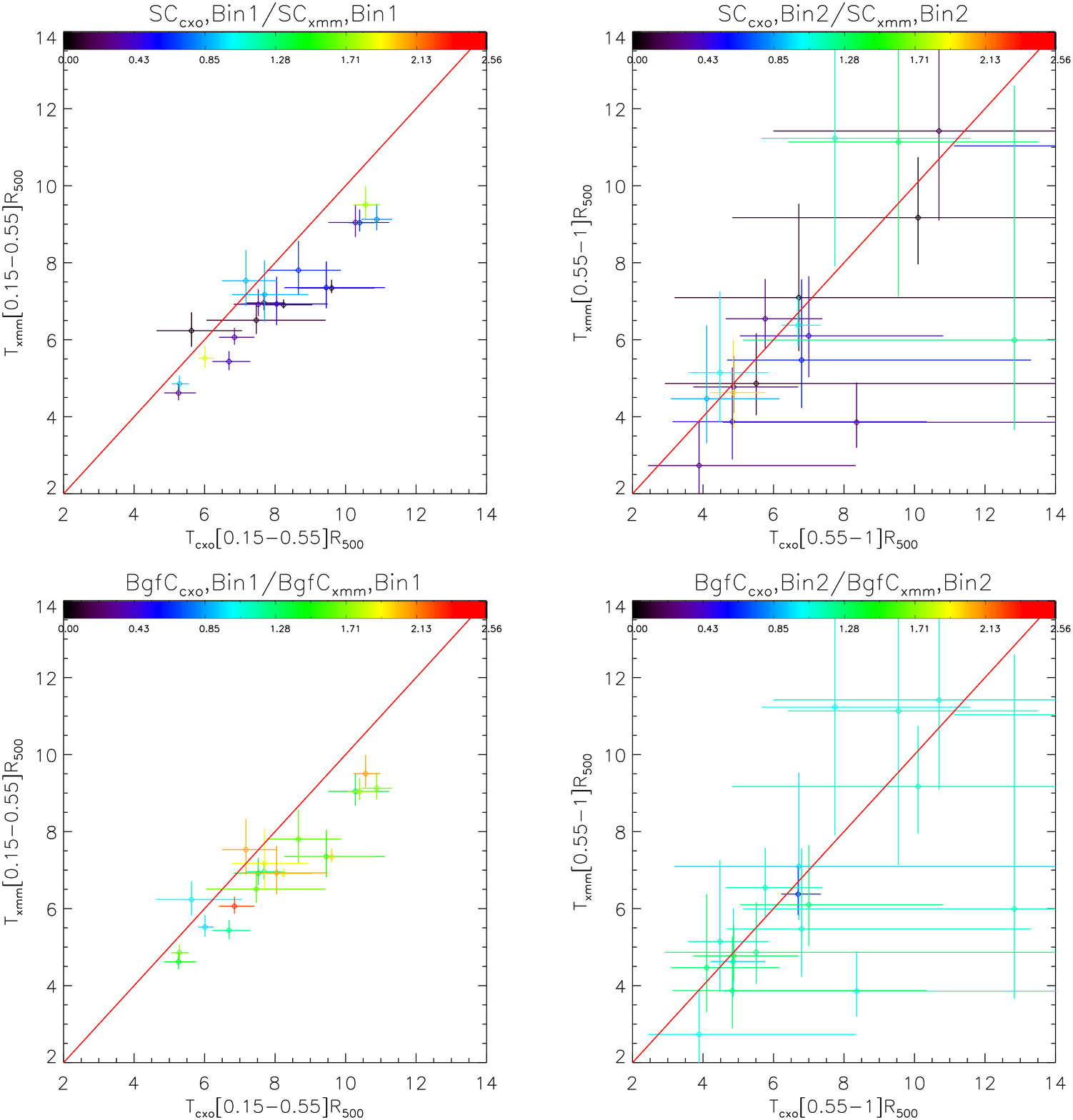}
    \caption{\textit{Upper panel:}  \emph{Chandra} temperature versus the  \emph{XMM-Newton} temperature in the inner bin [0.15-0.55]$r_{500}$ (left panel) and in the outer bin [0.15-0.55]$r_{500}$ (right panel). The clusters are colour-coded according to the ratio between the \emph{Chandra} and  \emph{XMM-Newton} net counts. The red line is the relation 1:1. \textit{Bottom panel:} as upper panel but with the color indicating the ratio between the \emph{Chandra} and  \emph{XMM-Newton} background fraction.}
    \label{fig:cxo_xmm_T_counts}
  \end{center}
\end{figure*}

Furthermore, for each cluster, we extract the spectra in two annuli with 
a width equal to [0.15-0.55]$r_{500}$ (Bin1) and  [0.55-1.0]$r_{500}$ (Bin2), defining for our subsample two regions close to $r_{2500}$ and $r_{500}$, respectively.
Our purpose is to investigate if, in each bin, there is any systematics in the temperature estimate due to the number of source counts ($SC_{cxo}$, Bin$i$ and $SC_{xmm}$, Bin$i$) or to the fraction of background events in the spectra ($BgfC_{cxo}$, Bin$i$, $BgfC_{xmm}$, Bin$i$).
In Figure~\ref{fig:cxo_xmm_T_counts}, in the upper panel, we plot the \emph{Chandra} temperature versus the  \emph{XMM-Newton} temperature in the inner bin [0.15-0.55]$r_{500}$ (left panel) and in the outer bin [0.15-0.55]$r_{500}$ (right panel). The clusters are colour-coded according to the ratio between the \emph{Chandra} and  \emph{XMM-Newton} net counts. The red line is the relation 1:1. As expected, in the inner bin, we find that the \emph{Chandra} temperature are in mean higher than the 
  \emph{XMM-Newton} ones and that the disagreement is worse going to higher temperature. However, we can notice that the most of the estimated temperatures  are higher than 6 keV, that is, as already observed in other works \citep{2008A&A...478..575K,2013ApJ...767..116M} a $thresholding$ value to make the difference between  \emph{Chandra} and  \emph{XMM-Newton} significant. In the outer bin (right panel), instead,  we can notice that the most of the estimated temperatures are $\leq$ 6 keV. Furthermore, larger errors bars are associated to highest temperature, due to the larger difficulty to distinguish the hottest spectra having a flatter shape from the background. For this reason,  these clusters have a \textit{lower weight} in the fit of the statistical sample.  However, as suggested by the colour distribution indicating the ratio between the \emph{Chandra} and  \emph{XMM-Newton}, the net counts number affects the error bars in the outer bin but there is not evident correlation of it with the global temperature (in both bins). 
 In the bottom panel of Figure~\ref{fig:cxo_xmm_T_counts} we plot the same quantities of the upper panel except for the color now indicating the ratio between the \emph{Chandra} and  \emph{XMM-Newton} background fraction. The typical background fraction is $0.18 \pm 0.04$ and $0.11 \pm 0.03$ in the inner bin and $0.75 \pm 0.16$ and $0.67 \pm 0.03$ in the outer bin for \emph{Chandra} and  \emph{XMM-Newton} respectively. We can notice that also in this case, there is no correlation between the temperature estimated in each bin and the number of counts attributed to the background.

We also explore mass dependence of, and intrinsic scatter in, the
relationship between \emph{Chandra}- and \emph{XMM-Newton}-based
masses, fitting a two parameter model: $M_{\rm CXO}=a\,M_{\rm
  XMM}^\alpha$ to the data using the Bayesian method described by
\cite{2007ApJ...665.1489K}.  The slope parameter $\alpha$ is
consistent with unity at all three over-densities for both gas masses
and hydrostatic masses (Table~\ref{tab:masscomp}).  The intrinsic
scatter is estimated to be $\sim3\%$ for gas masses and $\sim6-8\%$
for hydrostatic masses.

\begin{table}
  \caption{Comparison of gas mass measurements $M_{\rm
  CXO}/M_0=a_{\Delta}\,(M_{\rm XMM}/M_0)^{\alpha}$}
  
  \begin{tabular}{ccccc}
    \hline
    \noalign{\smallskip}
   
    $\Delta$ & $M_0$&\multispan1{\hfil Fixed-slope Model\hfil} &
    \multispan2{\hfil  Two parameters Model\hfil} \\
    & $M_\odot$& $a_\Delta$  & $a_\Delta$ & $\alpha$ \\
    \noalign{\smallskip}
    \hline
    \noalign{\smallskip}
    2500 &  $3\times10^{13}$ &  $0.97\pm0.02$ &   $0.99\pm0.02$ & $0.95\pm0.05$  \\
    1000 & $6\times10^{13}$ &   $0.98\pm0.02$ &   $0.98\pm0.02$ & $0.96\pm0.06$  \\
    500  & $9\times10^{13}$ &     $0.99\pm0.02$ &   $0.99\pm0.02$ &  $0.98\pm0.06$  \\
    \noalign{\smallskip}
        \hline
  \end{tabular}
  \label{tab:gas_mass_comp}
\end{table}

\begin{table}
  \caption{Comparison of hydrostatic mass measurements $M_{\rm
      CXO}/M_0=a_{\Delta}\,(M_{\rm XMM}/M_0)^{\alpha}$}
  \begin{tabular}{ccccc}
    \hline
    \noalign{\smallskip}
   
    $\Delta$ & $M_0$&\multispan1{\hfil Fixed-slope Model\hfil} &
    \multispan2{\hfil  Two parameters Model\hfil} \\
    & $M_\odot$& $a_\Delta$  & $a_\Delta$ & $\alpha$ \\
    \noalign{\smallskip}
    \hline
    \noalign{\smallskip}
    2500 &  $3\times10^{13}$ &  $1.15\pm0.05$ &   $1.15\pm0.05$ & $1.01\pm0.10$  \\
    1000 & $5\times10^{13}$ &   $1.06\pm0.05$ &   $1.08\pm0.06$ & $0.99\pm0.12$  \\
    500  & $7\times10^{13}$ &     $1.02\pm0.05$ &   $1.04\pm0.06$ &  $1.01\pm0.15$  \\
    \noalign{\smallskip}
        \hline
  \end{tabular}
  \label{tab:masscomp}
\end{table}

%%%%%%%%%%%%%%%%%%%%%%%%%%%%%%%%%
\subsection{X-ray/weak-lensing mass comparison}
\label{xray_wl_mass}
%%%%%%%%%%%%%%%%%%%%%%%%%%%%%%%%%%%
\begin{figure*}
  \begin{center} \leavevmode
    \includegraphics[width=1.\textwidth]{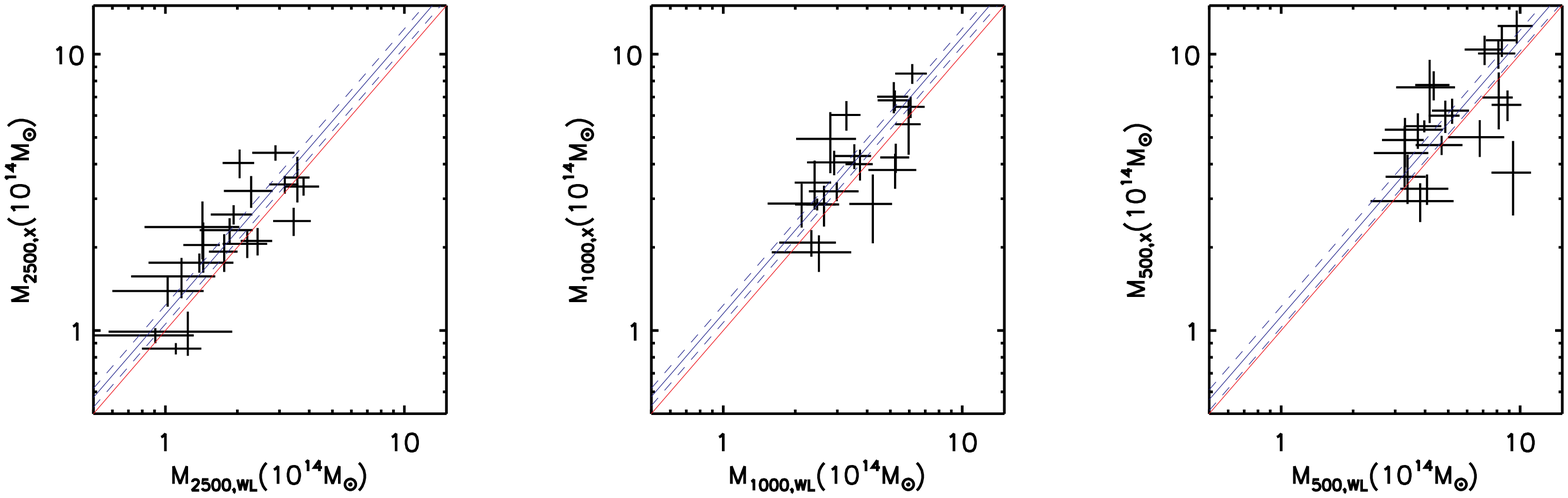}
    \caption{\textit{Left panel}: X-ray mass versus weak lensing mass  at
      $r_{2500}$: the solid blue line is the best fit relation with fixed slope = 1; the
      dashed blue lines are the errors of the best fit at $1\sigma$,
      the solid red line is the expected line 1:1. \textit{Middle
        panel}: X-ray mass versus weak lensing mass  at $r_{1000}$,
      similarly for X-ray versus weak lensing mass  at
      $r_{2500}$. \textit{Bottom panel}: X-ray mass versus weak lensing mass
      at $r_{500}$, with lines similar for $r_{2500}$ and
      $r_{1000}$.}
    \label{fig:fitmx_mwl}
  \end{center}
\end{figure*}

We now compare our hydrostatic mass measurements with published
  LoCuSS weak-lensing mass measurements
  \citep{2010PASJ...62..811O,2013ApJ...769L..35O}.
  \cite{2013ApJ...769L..35O} presented a stacked weak-lensing analysis
  of all 50 clusters in the High-$\Lx$ sample, showing that systematic
  errors are sub-dominant to statistical errors.  This was achieved in
  part via a conservative selection of background galaxies redder then
  the cluster red sequence.  The low number density of background
  galaxies therefore precluded analysis of individual clusters.
  Stacked weak-lensing analysis of the clusters in common between
  Okabe et al.\ (2010) and Okabe et al.\ (2013) revealed that on
  average the $M_{500}$ measurements from the latter exceed those from
  the former by 20\%.  In anticipation of a more careful analysis
  based on individual cluster measurement, we therefore compare our
  new hydrostatic mass measurements with weak-lensing masses from
  Okabe et al.\ (2010), applying a 20\% increase to the latter.  This
  comparison is applied to the clusters in sub-sample 3 (Table~1).

  We use hydrostatic masses from \emph{Chandra}, where available, and
  from \emph{XMM-Newton} otherwise.  All hydrostatic masses are
  calculated at the over-density radii derived from the 2010
  weak-lensing analysis.  We follow the same approach as in the
  previous section to measure the mean mass ratio: $M_{\rm
    X}=a_{\Delta}\,M_{\rm WL}$.  Our hydrostatic masses exceed
    the weak-lensing masses of \cite{2010PASJ...62..811O} by
    $\sim10-15\%$ at all over-densities with no obvious trend with
    over-density and dynamical state.  Specifically, at $r_{500}$ we find: $\Mx/\Mwl=1.12\pm0.10$.
    Therefore applying the 20\% increase to the
      \cite{2010PASJ...62..811O} weak-lensing masses discussed above
      suggests that at $r_{500}$ our X-ray and weak-lensing data are
      consistent with $\Mx/\Mwl\simeq0.93$.  This is in qualitative
      agreement with the theoretical expectation that $\Mx<\Mwl$, and
      is consistent with recent observational and theoretical studies
      \citep{2007ApJ...655...98N, 2008MNRAS.384.1567M,
        2010ApJ...711.1033Z,2013ApJ...767..116M}.

A more detailed comparison is possible between our hydrostatic
  mass measurements and the weak lensing mass of
\citet{2013ApJ...767..116M}.  There are 21 clusters in common
  between our samples (sub-sample 4 in Table~1).  Using
  \citeauthor{2013ApJ...767..116M}'s weak lensing mass and computing
  our hydrostatic mass measurements within their values of $r_{500}$,
we obtain $\Mx/\Mwl=1.07\pm0.06$ at $\Delta=500$ for these 21
clusters.  To achieve a careful like-for-like comparison, we
  repeat this calculation using Mahdavi et al.'s hydrostatic and
  weak-lensing measurements for the same sub-sample of 21 clusters,
  obtaining $\Mx/\Mwl=0.94\pm0.07$, which slightly exceeds the value
  of $\Mx/\Mwl=0.88\pm0.05$ that they obtain for their full sample.
  Our hydrostatic mass measurements at $\Delta=500$ therefore exceed
  Mahdavi et al.'s measurements by $\sim14\%$.  Interestingly, this
  excess matches the difference between Mahdavi et al.'s
  \emph{XMM-Newton}- and \emph{Chandra}-based mass measurements at
  $\Delta=2500$.  They derived an energy-dependent correction to the
  \emph{Chandra} effective area based on their cross-calibration at
  $\Delta=2500$, which acts to reduce their \emph{Chandra}-based mass
  measurements.  This correction also reduces their \emph{Chandra}
  mass measurements at $\Delta=500$ (priv.\ comm.\ A.\ Mahdavi).  We
  also find that the satellites disagree at $\Delta=2500$ (\S6.1),
  however we achieve close agreement between them at $\Delta=500$,
  without invoking any correction to the \emph{Chandra} effective
  area.  We suggest that the difference between our respective
  hydrostatic mass measurements at $\Delta=500$ can be in part
  explained by these differences between our analyses.  We also note
  that our hydrostatic mass measurements imply that Mahdavi et al.'s
  weak-lensing masses may be biased low.

%%%%%%%%%%%%%%%%%%%%%%%%%%%%%%%%%%%%%%%%%%%%%%%%%%%%%
\subsection{Scaling relation between hydrostatic mass and integrated
  $Y$ parameter}  \label{xray_sz}
%%%%%%%%%%%%%%%%%%%%%%%%%%%%%%%%%%%%%%%%%%%%%%%%%%%%%%

We compare our X-ray hydrostatic masses and the SZ signal,
$Y_{\rm sph}D_A^2$, measured with the SZA, an 8-element radio interferometer optimized for measurements of the SZ effect  \citep{2012ApJ...754..119M},
for the 17 clusters in sub-sample~2 (Table~1).  Following
\citeauthor{2012ApJ...754..119M}, we fit the following model to the
data: \be \frac{M_{X}(r_{\Delta})}{10^{14}M_{\odot}}= 10^{A} \left(
\frac{Y_{sph}D_{A}^{2}E(z)^{-2/3}}{10^{-5}\rm{Mpc}^{2}}\right)^{B},
\label{scaling_rel}
\ee performing the regression in linearized coordinates using the
base-10 logarithm of the data points, and using the Bayesian
regression method described by \cite{2007ApJ...665.1489K}.  We
  perform the fits twice at each over-density, once with both $A$,
  $B$, and the intrinsic scatter as free parameters, and once with the
  slope parameter $B$ fixed at the self-similar value of $B=3/5$.

Our best-fit slopes and normalizations of the fits are consistent
within $68\%$ confidence intervals with the $M_{\rm WL}-\Ysph$
relation of \citeauthor{2012ApJ...754..119M} (Table~\ref{tab:m_Ysz}).
The fit with a free slope parameter is also consistent with the
self-similar slope.  Formally, the data are consistent with
zero intrinsic scatter in mass at fixed $\Ysph$ around the best-fit
scaling relation (Table~\ref{tab:m_Ysz}).  However, despite the large
uncertainties in our measurement of the intrinsic scatter,
we note that the scatter of $\sim15-20\%$ is similar to that obtained in other studies
\citep[e.g.][]{2006ApJ...650..538N,2012ApJ...754..119M}, and appears
to decrease toward larger radii.

\begin{figure*} 
  {\centering \leavevmode
    \includegraphics[width=1.1\textwidth]{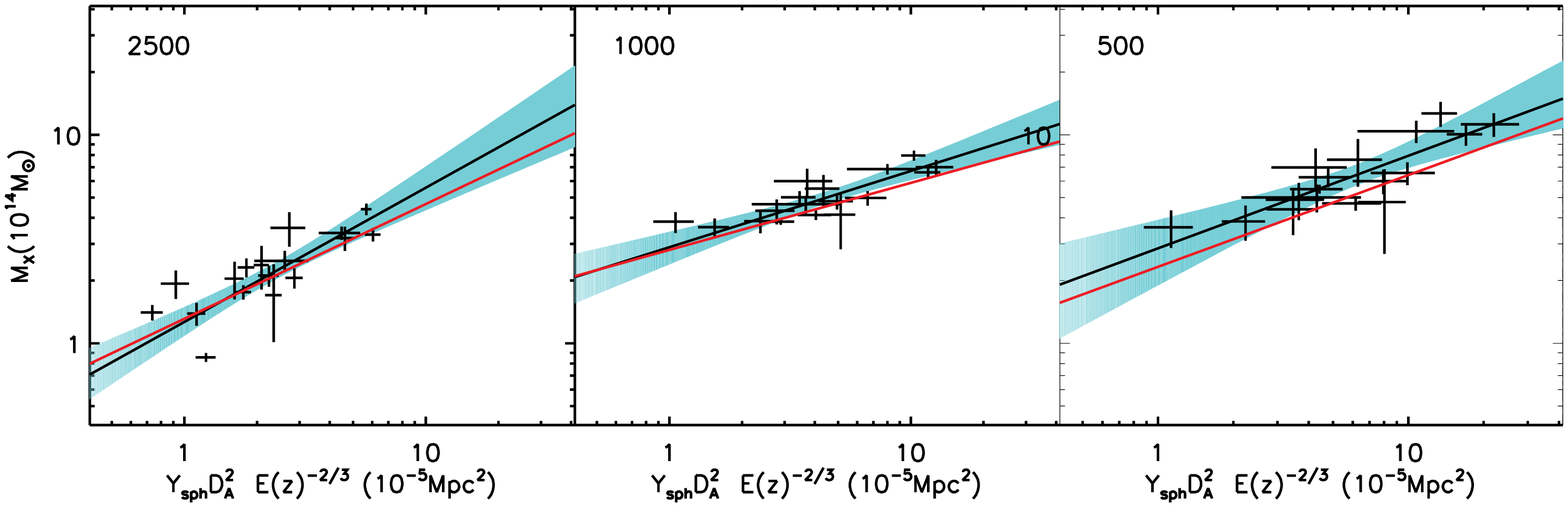} }
  \caption{Our best-fit $M_{\rm X,\Delta}-\Ysph\,D_{A}^{2}$ relations
    (blue line with cyan error envelope) at $\Delta=$2500 (left), 1000
    (middle) and 500 (right), plus the best-fit $M_{\rm WL}-\Ysph$
    relations from \protect\cite[][red]{2012ApJ...754..119M}.}
  \label{fig:scal_relat_M_sz_sub}
\end{figure*}

\begin{figure} 
  {\centering \leavevmode
    \includegraphics[width=.5\textwidth]{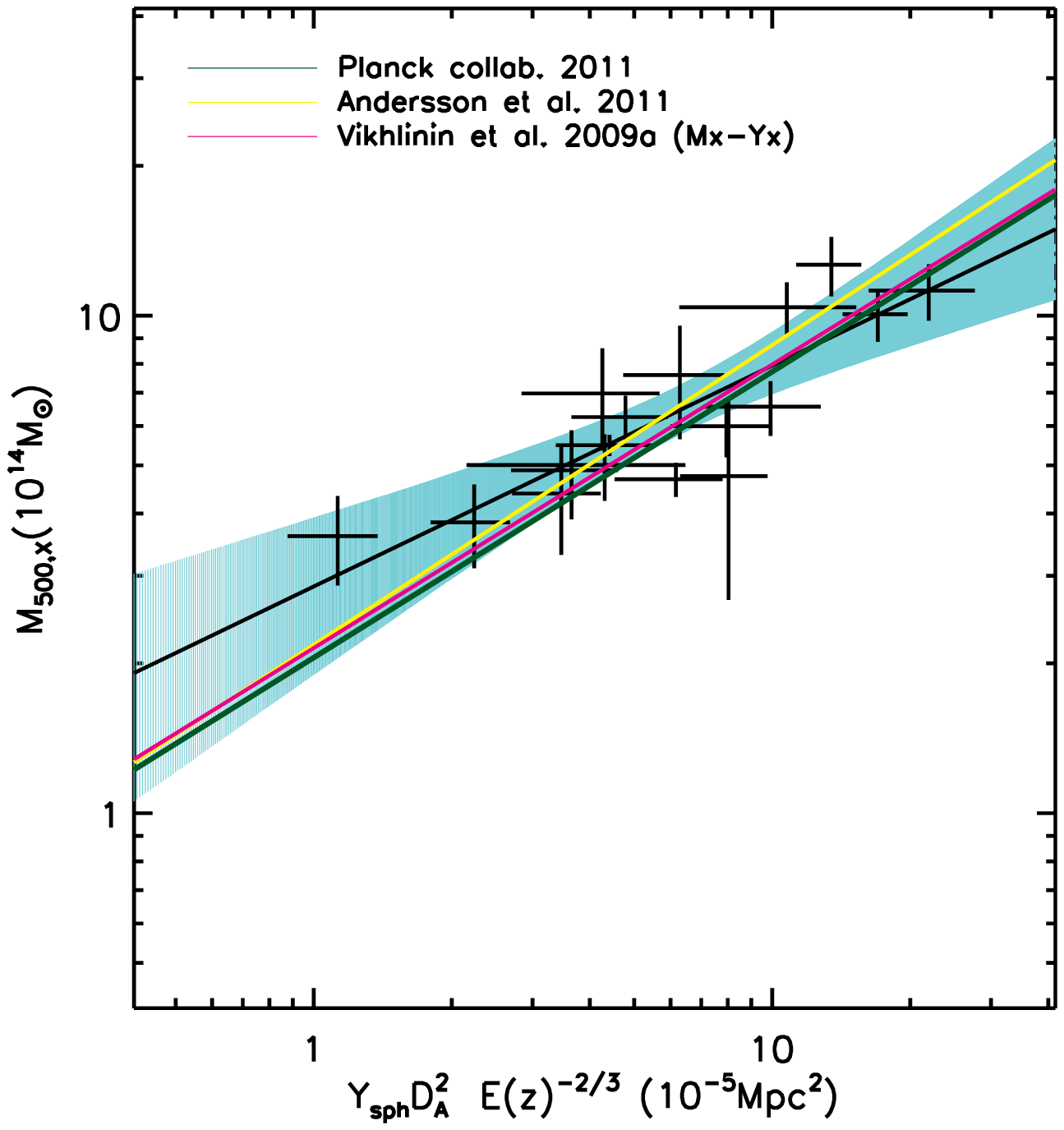}}
  \caption{Our best-fit $M_{\rm X,\Delta}-\Ysph\,D_{A}^{2}$ relations
    (blue line with cyan error envelope) at $\Delta=$500, plus  the
    best-fit relations from
    \protect\cite[][green]{2011A&A...536A..11P} and
    \protect\cite[][yellow]{2011ApJ...738...48A}. In magenta is
    plotted the best-fit $M_{\rm X}-Y_{\rm X}$ relation from
    \protect\cite{2009ApJ...692.1033V}}
  \label{fig:scal_relat_M_sz_sub500}
\end{figure}

At $\Delta=500$ (Fig~\ref{fig:scal_relat_M_sz_sub500}), our results agree with
\citet[yellow line][]{2011ApJ...738...48A}, who estimated cluster mass using the
observed $M_{\rm X,500}-Y_{\rm X}$ scaling relation
\citep{2009ApJ...692.1033V}. Furthermore, in the same figure, we
 compare our best-fit $M_{\rm X,500}-\Ysph\,D_{A}^{2}$ relation with the observed $M_{\rm X,500}-\Ysph\,D_{A}^{2}$ relation obtained by \cite[green line][]{2011A&A...536A..11P} using 
$\Ysph$ measured by the {\it Planck} satellite and  mass $M _{\rm X,500}$ estimated from the $M_{500}-Y_{X,500}$ relation given in \cite{2010A&A...517A..92A}.
However we can notice in our relation that the cluster ABELL0383  affects the slope far more than any other. As discussed also in other works \citep{2012ApJ...754..119M, 2012arXiv1204.2743P}, this cluster shows unusually low SZ flux for its apparent mass in their SZA observations. In fact, even if ABELL0383 seems to be a very relaxed system in X-rays, \cite{2012MNRAS.420.1621Z} find that it is a cluster- cluster lens system, having at least two other well-defined optical structures within 15'.
 
Furthermore,
the $\Mx-\Yx$ relation obtained by
  \cite{2009ApJ...692.1033V} (magenta line in
  Fig~\ref{fig:scal_relat_M_sz_sub500}) is consistent with our $\Mx-\Ysph$ relation, after
  re-normalising $Y_{\rm X}$ by the factor $C_{XSZ}$ following
  \cite{2012arXiv1204.2743P}.  This is consistent with the ratio
  between $\Ysph$ and $C_{XSZ}Y_{\rm X}$ being close to unity, in
  agreement with \cite{2012arXiv1204.2743P}, who obtained
  $\Ysph/(C_{XSZ}Y_{\rm X})=0.95\pm0.04$.  We will investigate the
  $\Ysph-(C_{XSZ}Y_{\rm X})$ relationship for our full sample in a
  future paper.

\begin{table}
  \begin{center}
    \caption[]{ $M_{\rm X}-Y_{\rm{sph}}$ scaling relation\\}
    % \resize
    \begin{tabular}{lccc}
      \hline
        \noalign{\smallskip} 

      $\Delta$    & $A$ & $B$ &  $\sigma_{M|Y}$  \\ 
        \noalign{\smallskip} 

      \hline
        \noalign{\smallskip} 

      2500        & $0.101\pm0.070$ & $0.649\pm0.153$ & $0.225^{+0.111}_{-0.083}$  \\
       \noalign{\smallskip}
      1000	  & $0.273\pm0.114$ & $0.547\pm0.159$ & $0.153^{+0.116}_{-0.071}$ \\
       \noalign{\smallskip}
      500         & $0.454\pm0.154$ & $0.446\pm0.182$ & $0.157^{+0.127}_{-0.073}$ \\  
       \noalign{\smallskip}  
      \hline
        \noalign{\smallskip} 

            2500        & $0.117\pm0.038$ &3/5& $0.215^{+0.109}_{-0.120}$  \\
             \noalign{\smallskip}
      1000	  & $0.213\pm0.041$ & 3/5 & $0.136^{+0.106}_{-0.065}$ \\
       \noalign{\smallskip}
      500         & $0.303\pm0.054$ & 3/5 & $0.149^{+0.121}_{-0.069}$ \\   
       \noalign{\smallskip} 

      \hline
    \end{tabular}
    \label{tab:m_Ysz}
  \end{center}
\end{table}

%%%%%%%%%%%%%%%%%%%%%%%%%%%%%%%%%%%%%%%%%%%%%%%%%%%%%%%%%%%%%%%%%
%							 CONCLUSION
%%%%%%%%%%%%%%%%%%%%%%%%%%%%%%%%%%%%%%%%%%%%%%%%%%%%%%%%%%%%%%%%%%%%%%%%

\section{Discussion}\label{sec:discuss}

A primary goal of comparing hydrostatic mass measurements from X-ray
observations to weak-lensing mass measurements is to test the
assumption that the cluster gas is in hydrostatic equilibrium.
Predictions derived from cosmological numerical simulations indicate
that the most prominent departures from hydrostatic equilibrium occur
at large radius, $r\gs r_{500}$, where bulk motion of gas provides
additional, non-thermal pressure support, biasing the hydrostatic
masses low \citep[e.g.][]{2009ApJ...705.1129L, 2010ApJ...725...91B,
  2006MNRAS.369.2013R,2007ApJ...655...98N, 2010A&A...514A..93M}.

In observationally establishing the magnitude of this bias, the
systematic errors in the mass measurements must be accounted for.
X-ray mass measurements may be systematically biased by uncertainties
in instrumental calibration
\citep{2010A&A...523A..22N,2011A&A...525A..25T}, and differences in
analysis methods \cite[e.g.][]{2011arXiv1106.4052M}.  It can be
difficult to disentangle these two effects, with up to $\sim45\%$
differences between mass measurements based on independent
observations and analysis of the same clusters reported in some cases
\citep{2012arXiv1204.6301R}.  Applications of a single analysis method
to both \emph{Chandra} and \emph{XMM-Newton} data tend to report
smaller systematic differences in mass measurements, for example,
\citet{2013ApJ...767..116M} recently reported a $15\%$ systematic
excess of \emph{Chandra}-based $M_{500}$ measurements over
\emph{XMM}-based measurements of the same.

Applying a consistent method to identical cluster regions for both
satellites, we find \emph{Chandra} and \emph{XMM-Newton} hydrostatic
mass measurements that agree very well. In particular, at $r_{500}$ we
obtain $M_{\rm CXO}/M_{\rm XMM}=1.01\pm0.05$.  This agreement is
achieved by combining the latest instrumental calibration with our new
analysis method, incorporating an analytic model of the X-ray
background.  Our hydrostatic mass measurements are also consistent
with a $\sim10\%$ non-thermal pressure support at $r_{500}$, based on
a comparison between our measurements and weak-lensing mass estimates
of individual clusters from Okabe et al.\ (2010) and from the stacked
analysis of Okabe et al.\ (2013).  We will investigate non-thermal
pressure support in more detail in a future article.

\section{Conclusion}

Accurate measurements of galaxy cluster masses are required to
constrain adequately the cosmological parameters as one of several
complementary probes.  In this paper, we present the total hydrostatic
mass and gas profiles for the 50 clusters in the approximately
mass-selected LoCuSS High-$\Lx$ sample, using all the available
\emph{Chandra} and \emph{XMM-Newton} observations.  These measurements
are performed using exactly the same procedure and also using two
newly developed analytic background models, one for \emph{Chandra}
ACIS-I and one for \emph{XMM-Newton}, that model the
spatial variation of the background with an accuracy better than 2\%
and than 5\%, respectively.  We use this sample to investigate the
cross-calibration of cluster mass measurements between \emph{Chandra}
and \emph{XMM-Newton}.  We also compare our X-ray mass with weak
lensing mass estimates and integrated Compton effect $Y_{SZ}$
measurements, investigating their observational relations and
comparing with numerical simulation predictions.  The main results are
summarized below.

\begin{enumerate}

\item For a subsample of 21 clusters, having both \emph{Chandra}
  ACIS-I and \emph{XMM-Newton} data, we derived the mass profiles
  extracting the surface brightness and the temperature profiles
  exactly in the same sky regions.  This allowed us to measure the
  cross calibration uncertainties among the two instruments.  We find
  that, for each cluster, the gas and total mass profiles are fully
  consistent within $1\sigma$. In particular, the average ratios
  between the \emph{Chandra} and \emph{XMM-Newton} gas mass, $M_{\rm
    gas,CXO}/M_{\rm gas,XMM}$, are $0.98\pm0.02$ and $0.99,\pm0.02$ at
  $r_{1000}$, and $r_{500}$, respectively.  Agreement between
    the total hydrostatic masses improves to larger radii, with
    excellent agreement at $r_{500}$: $M_{\rm CXO}/M_{\rm
      XMM}=1.02\pm0.05$.  We find no evidence of the mass ratio being
    a function of mass, and just $\sim7\%$ intrinsic scatter.

\item For a subsample of 22 clusters, we compare hydrostatic
  masses with published weak-lensing mass estimates from
  \cite{2010PASJ...62..811O,2013ApJ...769L..35O}.  We find that the
  X-ray and weak-lensing mass measurements \cite{2013ApJ...769L..35O} are consistent with
  non-thermal pressure support of $\sim7\%$ at $r_{500}$.  Whilst this
  is consistent with other recent studies, we caution that further
  careful analysis of individual cluster mass measurements from our
  weak-lensing observations are required before firm conclusions can
  be drawn.

\item For a subsample of 17 clusters, we investigate the scaling
  relation between our X-ray hydrostatic masses and integrated
  Compton parameter $Y_{sph}$ obtained with the Sunyaev-Zel'dovich
  Array presented by \cite{2012ApJ...754..119M}. We find that the
  $M_{X}-Y_{sph}$ scaling relations, measured at $\Delta=2500$, 1000
  and 500, have, on average, a normalization higher than
  \cite{2012ApJ...754..119M}, but consistent within the
  errors. Furthermore, we find a slope that is better reproduced by
  self similar models. Our results are also consistent at $r_{500}$
  with \cite{2011ApJ...738...48A} and \cite{2010A&A...517A..92A}.

\item For the $M_{X}-Y_{sph}$ scaling relations, we find an intrinsic
  scatter of $\sim15-22\%$, with the smaller scatter seen at larger radii.
  These values are slightly lower, but statistically consistent with
  \citeauthor{2012ApJ...754..119M}, who obtained an intrinsic scatter
  of $\sim 20\%$ at all radii.  At larger radii, our estimate of
  intrinsic scatter is also consistent with the $10-15\%$ predicted in
  numerical simulations \citep[e.g.][]{2006ApJ...650..538N}.

\end{enumerate}

In summary, we have presented an important step forward in our ability
to measure the hydrostatic mass of galaxy clusters free from
instrument- and background-related systematic errors.  In the future we
will compare our hydrostatic mass estimates with weak lensing masses
and integrated Compton parameter measurements for the full High-$\Lx$
LoCuSS sample.  This comparison, based on our \emph{full sample}, will
allow us to investigate robustly the shape, normalization and scatter
of key cosmological scaling relations at low redshift.

\section*{Acknowledgments}

We acknowledge financial contribution from contracts ASI-INAF
I/009/10/0, ASI-INAF I/088/06/0, PRIN-INAF-2009 grant ``Weighing
Galaxy Clusters With Strong and Weak Lensing''. PM acknowledges
financial support by grants GO2-13153X and AR2-13012X issued by the
Chandra X-ray Observatory Center.  GPS acknowledges support from the
Royal Society.  
We thank the anonymous referee for his/her thorough review and highly appreciate the comments and suggestions, 
which significantly contributed to improving the quality of the publication.
We thank Arif Babul and Gabriel Pratt  for comments on the draft
manuscript and Andisheh Mahdavi for helpful discussions.
We also acknlowdge Y.-Y. Zhang, the PI of ten XMM-Newton 
observations that were used in our study.

\appendix

\section{Effects of instrumental cross calibration on temperature
  estimates}

\begin{figure*}
  \begin{center} \leavevmode
 
    \includegraphics[width=.4\textwidth]{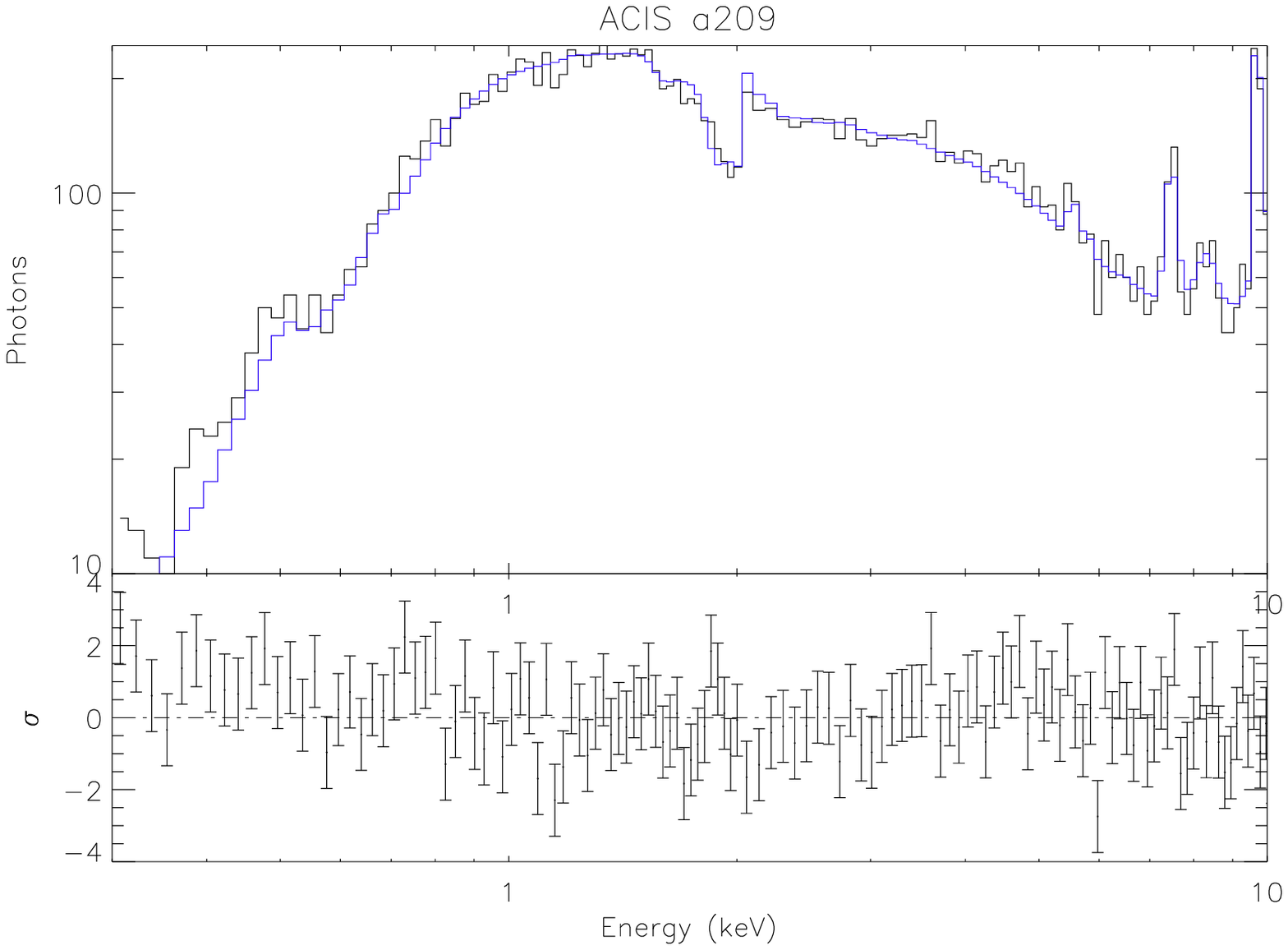}
    \includegraphics[width=.4\textwidth]{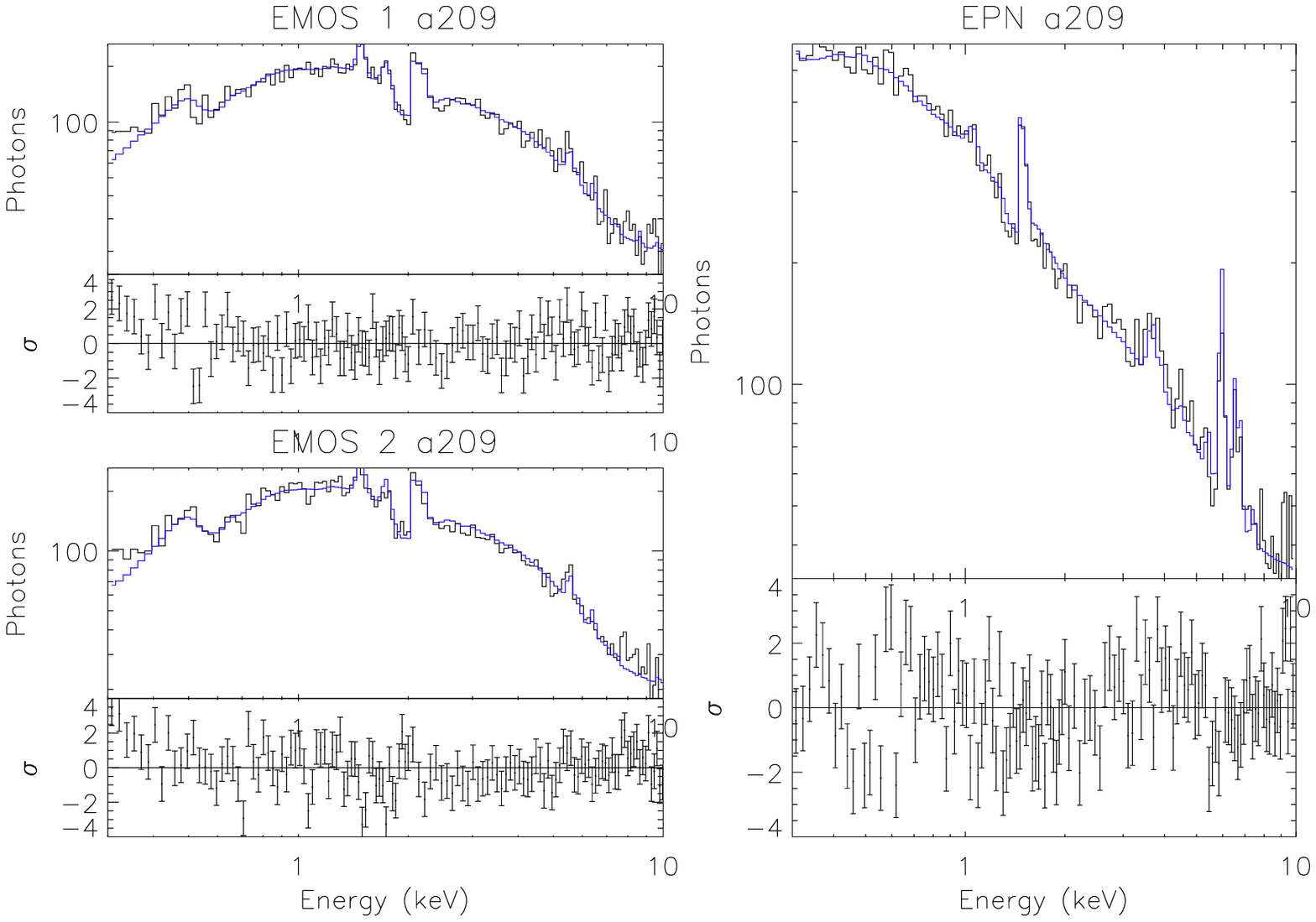}\\
    \includegraphics[width=.4\textwidth]{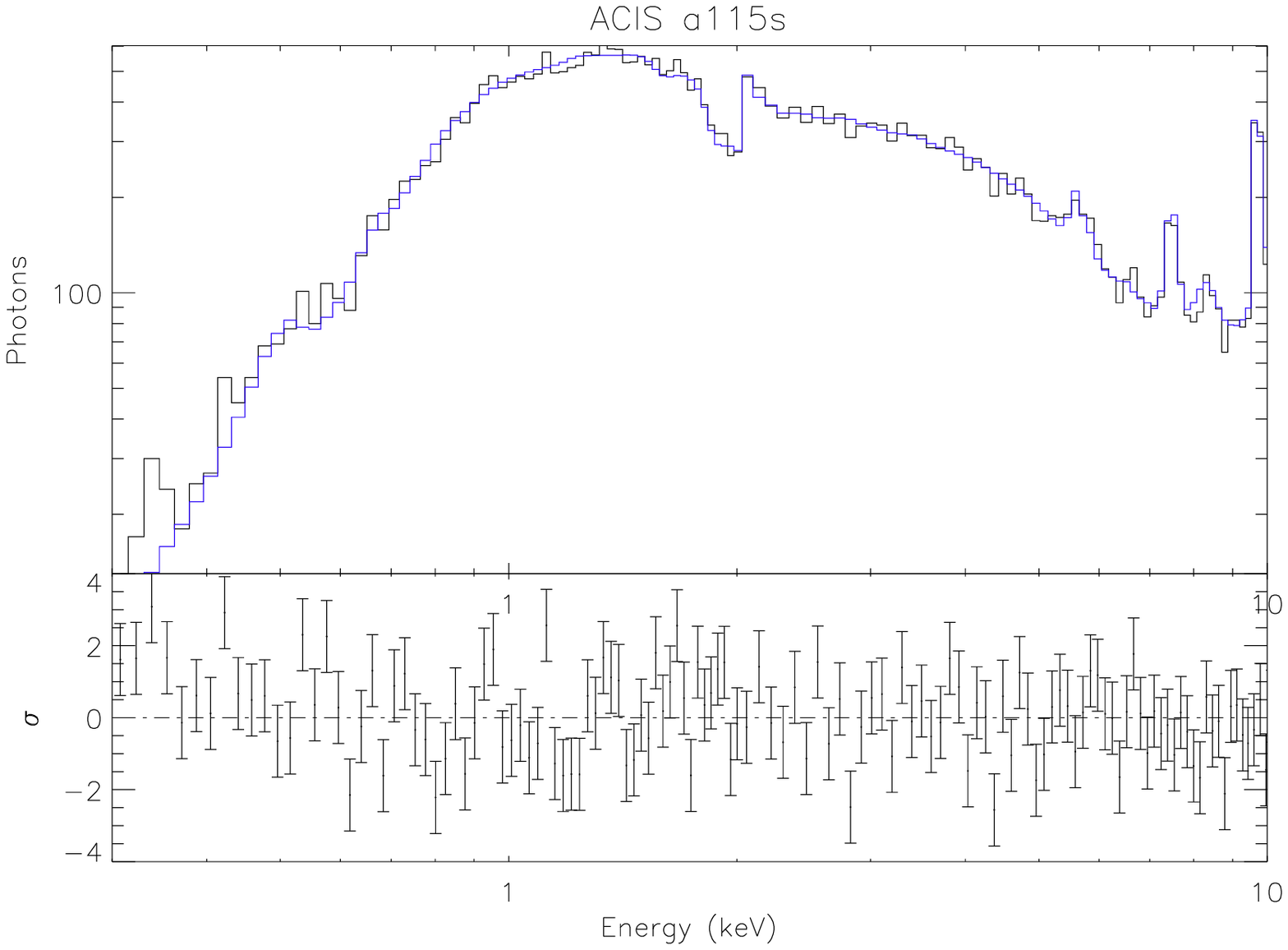}
    \includegraphics[width=.4\textwidth]{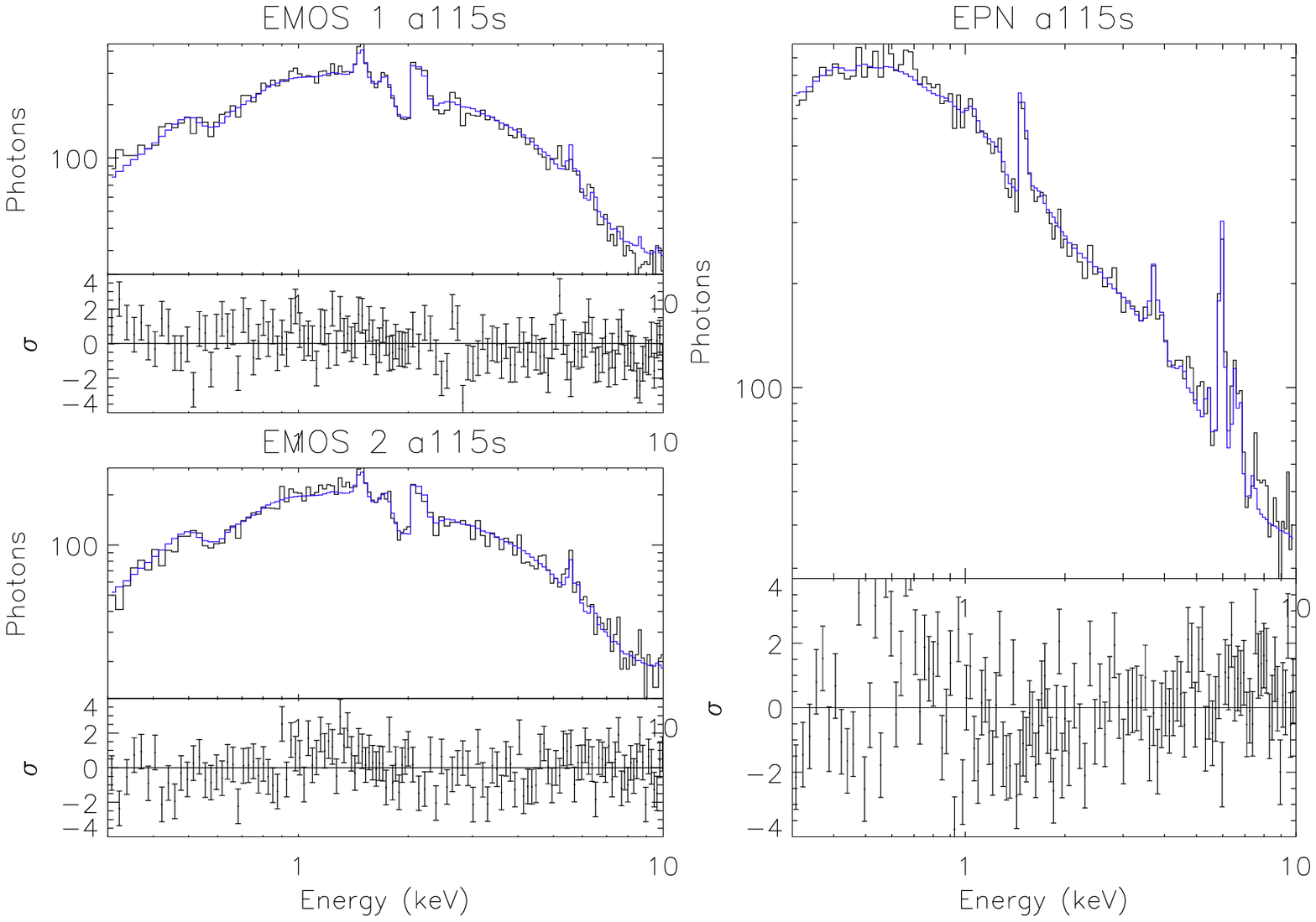}\\
    \includegraphics[width=.4\textwidth]{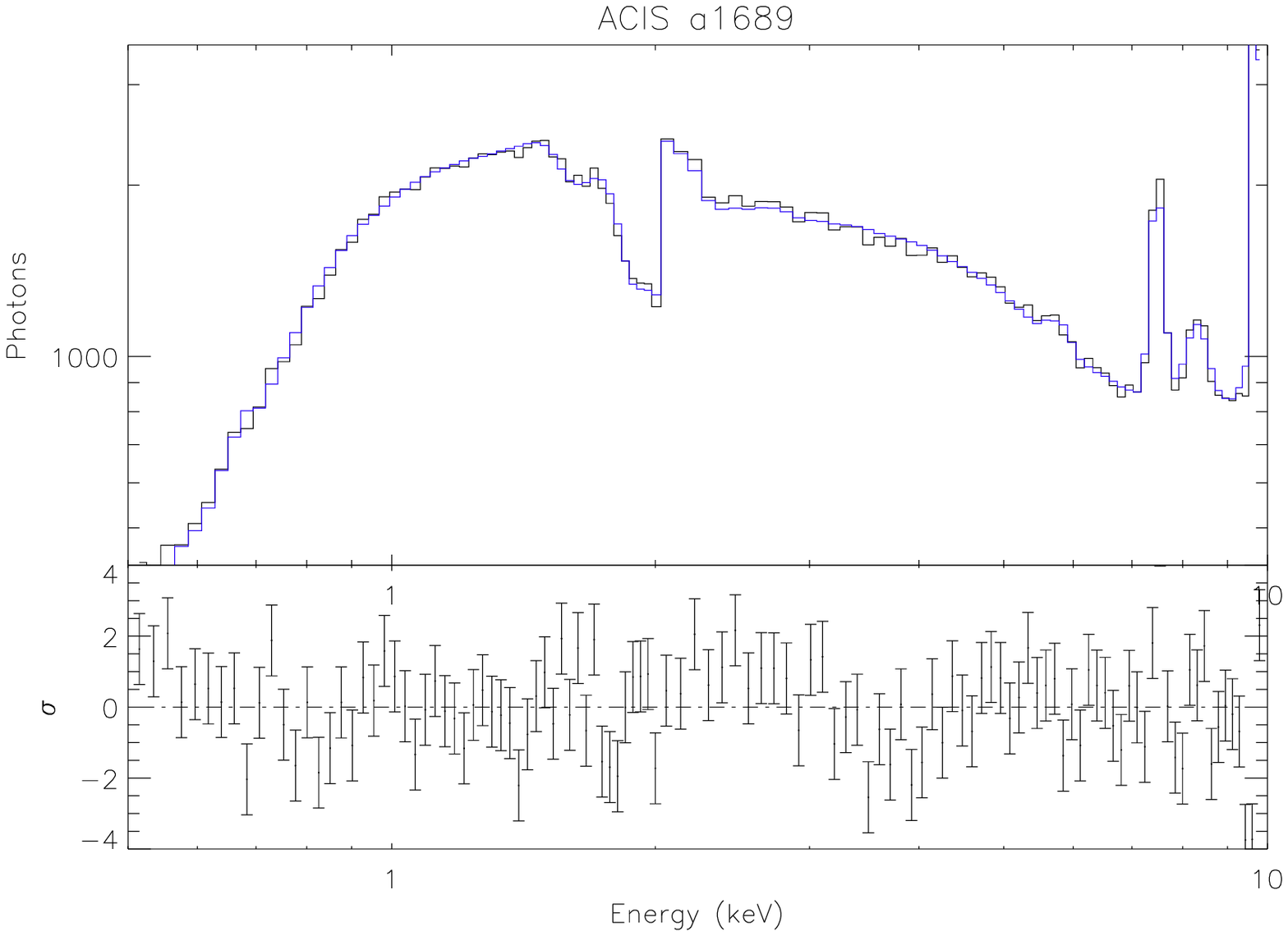}
    \includegraphics[width=.4\textwidth]{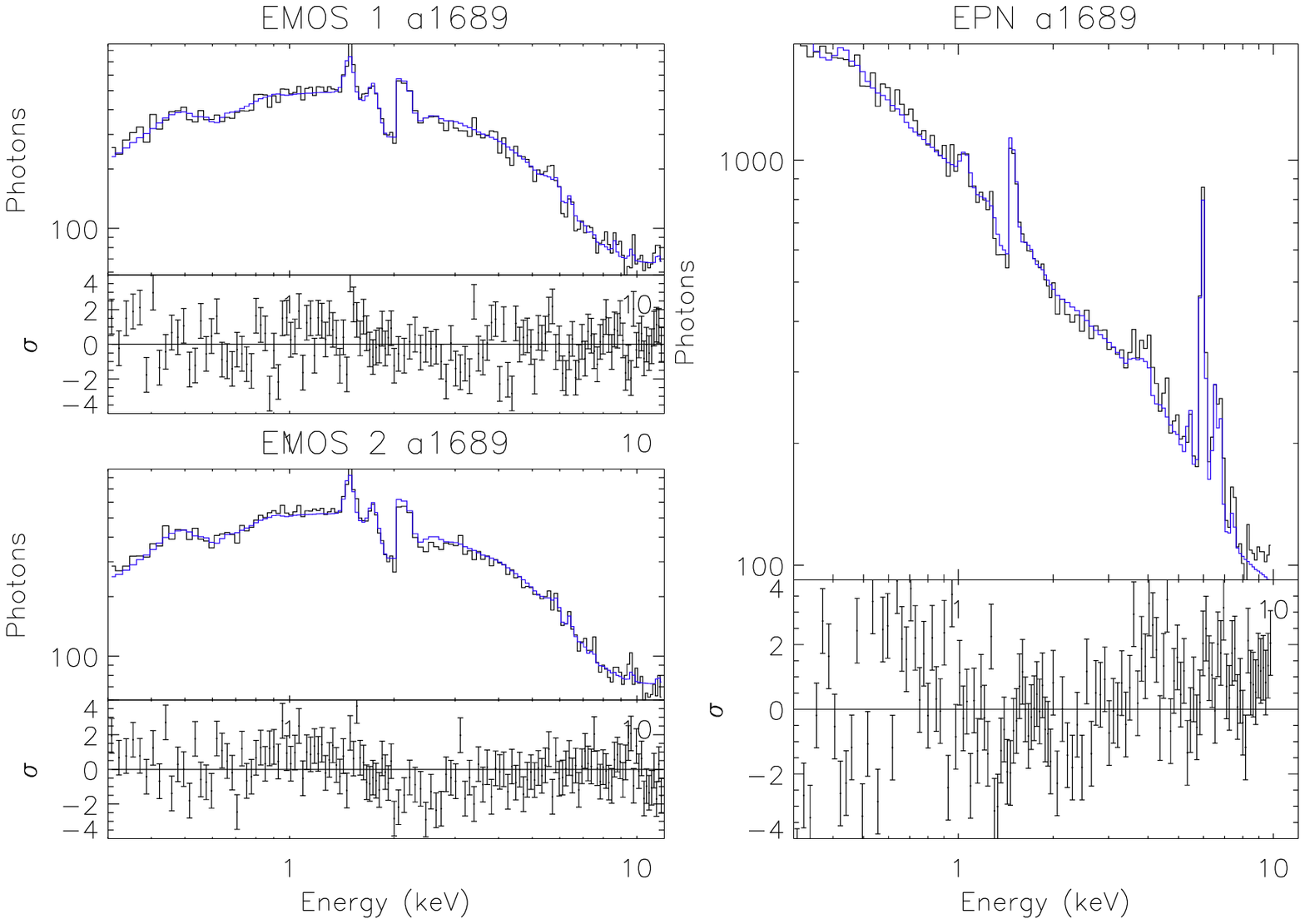}\\
    \caption{\textit{Top panel:} Spectra for the cluster a209 obtained
      by \emph{Chandra} ACIS-I and each of the \emph{XMM-Newton}
      cameras (MOS1, MOS2 and EPN). The \emph{Chandra} ACIS-I spectrum
      is fitted with an APEC model (blue line): the same model is
      overplotted on the spectra of the other \emph{XMM-Newton}
      cameras. Residuals are plotted in the lower portion of each
      panel.  Notice the good agreement between the data and the model
      for all four cameras within a 2$\sigma$ average confidence
      interval. \textit{Middle panel:} Same as top panel but for
      a115. The \emph{Chandra} best fit model agrees well with MOS1
      and MOS2 within 2$\sigma$ but not with PN.  \textit{Bottom
        panel:} Same as top panel but for a1689.  The \emph{Chandra}
      ACIS-I best fit model does not reproduce any of the spectra from
      the \emph{XMM-Newton} cameras, with differences beyond
      4$\sigma$. Furthermore, the spectra from the three
      \emph{XMM-Newton} cameras show significant differences among
      themselves.}
    \label{fig:calibrat_spectra}
  \end{center}
\end{figure*}

In \S\ref{gas_mass_cxo_xmm}, we show the existence of a good agreement
between the gas mass derived with \emph{Chandra} and \emph{XMM-Newton}
observations. This indicates that, on average, the flux measured by
\emph{Chandra} and \emph{XMM-Newton} in the [0.5-2.5] keV band is
consistent. Nevertheless, the \emph{Chandra} temperatures tend to be,
on average, higher than the \emph{XMM-Newton} ones, indicating that
there are still some cross calibration problems between the two
instruments.  Using the overlapping \emph{Chandra}-\emph{XMM-Newton}
sub-sample we quantify the total effect on the temperature based on
current calibration files.  This allow us to provide a lower limit for
the systematic uncertainties that affect the hydrostatic mass
measurements with \emph{Chandra} and \emph{XMM-Newton}, based on the
current calibration files.  To do this, for each cluster in the
subsample, we extract 4 spectra from the same circular region with
$r=r_{500}$ for ACIS-I, MOS1, MOS2 and PN.  In extracting the spectra,
we take care to exclude the common region which corresponds to the
brightest point sources present in the field of view plus the
combination of all the gaps of the four detectors.  To highlight
possible calibration problems, we consider the \emph{Chandra}
spectrum, which for convenience we use as reference spectrum for the
comparison, and fit it in the [0.3-10.0] keV energy band with an
absorbed APEC model with temperature, metallicity and $N_H$ as free
parameters. Then, we compare the resulting best fit model with the
spectra obtained by the other cameras, using the same normalization factor. We identify three different
cases that we show in the three panels of
Figure~\ref{fig:calibrat_spectra}.  In the first case, shown in the top
panel of the Figure~\ref{fig:calibrat_spectra}, we find that the best
fit model obtained by ACIS-I camera reproduces very well the spectra
of all the other \emph{XMM-Newton} cameras within a confidence level
better than $2\sigma$. It is worth noticing that this case seems to
hold for most of the clusters in our sample.

In the second case, shown in the middle panel in
Figure~\ref{fig:calibrat_spectra}, we find that the best fit model
obtained by ACIS-I camera reproduces well the spectra of some of the
cameras of \emph{XMM-Newton} but not all.  In the third case, shown in
the bottom panel in Figure~\ref{fig:calibrat_spectra}, we find that the
best fit model obtained by ACIS-I camera does not reproduce any of the
spectra from the cameras of \emph{XMM-Newton}.  We find a disagreement
between the \emph{Chandra} spectrum and each of the three
\emph{XMM-Newton} spectra at energy lower than 0.7 KeV, which may be
related to some bias in the modelling of the \emph{Chandra}
response. Such low energy discrepencies do not affect our overall
analysis, which is performed in the [0.7-10.]  KeV energy
band. Furthermore, the EPIC-PN spectra exhibit an excess at high
energy with respect to the ACIS-I and EPIC-MOS spectra. This could
partly be due to the presence of non-standard emission lines related
to solar flare contamination that are not predicted in the particle
background model. The EPIC-PN data has not been used for spectroscopic
measurements in the few systems of this kind in our sample.
 
\begin{figure*}
  \begin{center} \leavevmode
    \includegraphics[width=1.\textwidth]{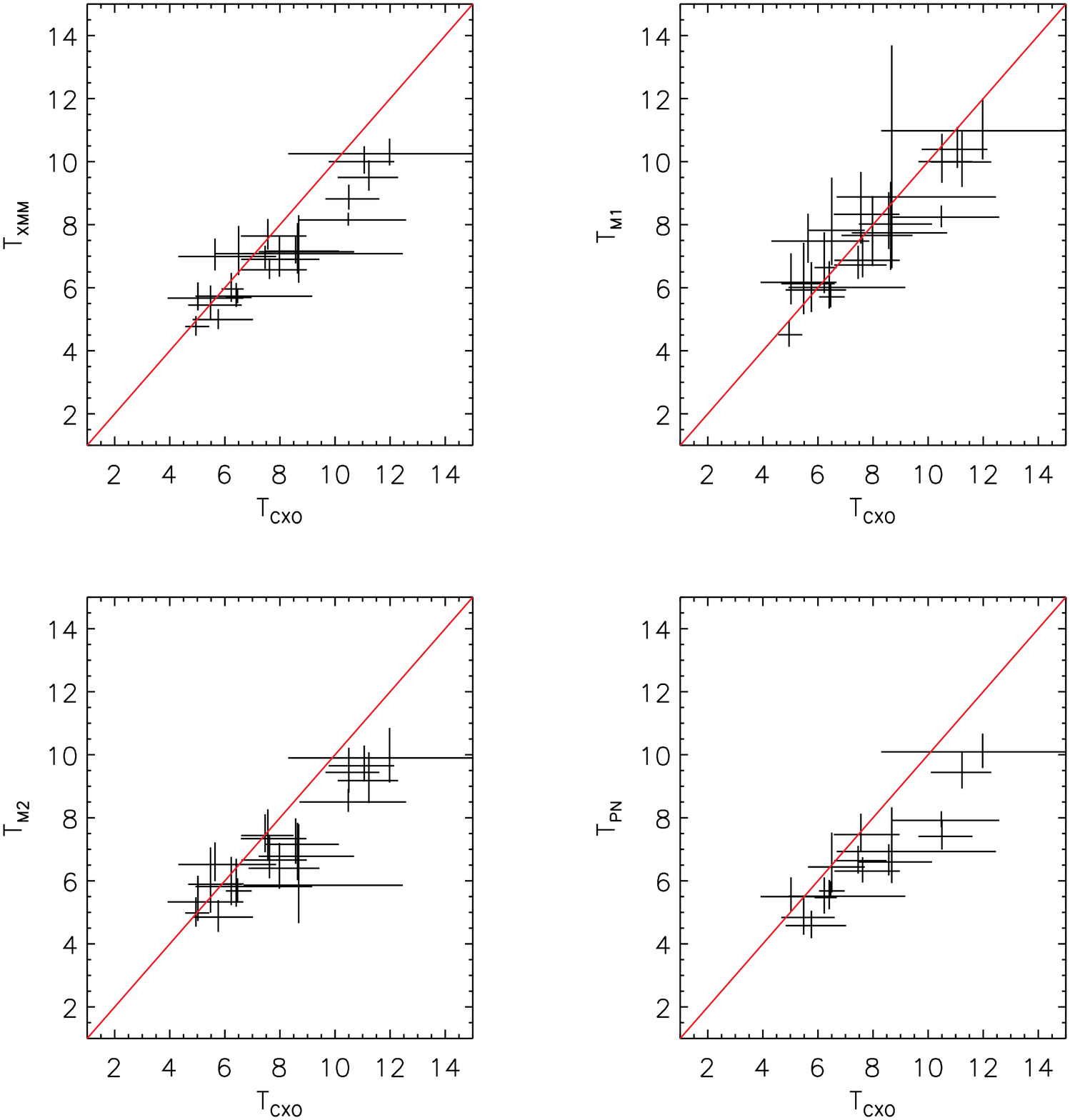}
    \caption{Comparison of the temperatures derived \emph{Chandra}
      with those derived from the various cameras of
      \emph{XMM-Newton}. The solid red line is the 1:1 relation.
      \textit{Top left panel}:Average of temperatures derived from all
      three \emph{XMM-Newton} cameras versus \emph{Chandra}
      temperatures.  \textit{Top right panel}: \emph{XMM-Newton} MOS1
      temperatures versus \emph{Chandra} temperatures.  \textit{Bottom
        left panel}: MOS2 versus \emph{Chandra}.  \textit{Bottom right
        panel}: EPN versus \emph{Chandra}}.
    \label{fig:cxo_xmm_temp}
  \end{center}
\end{figure*}

\begin{figure*}
  \begin{center} \leavevmode
    \includegraphics[width=1.\textwidth]{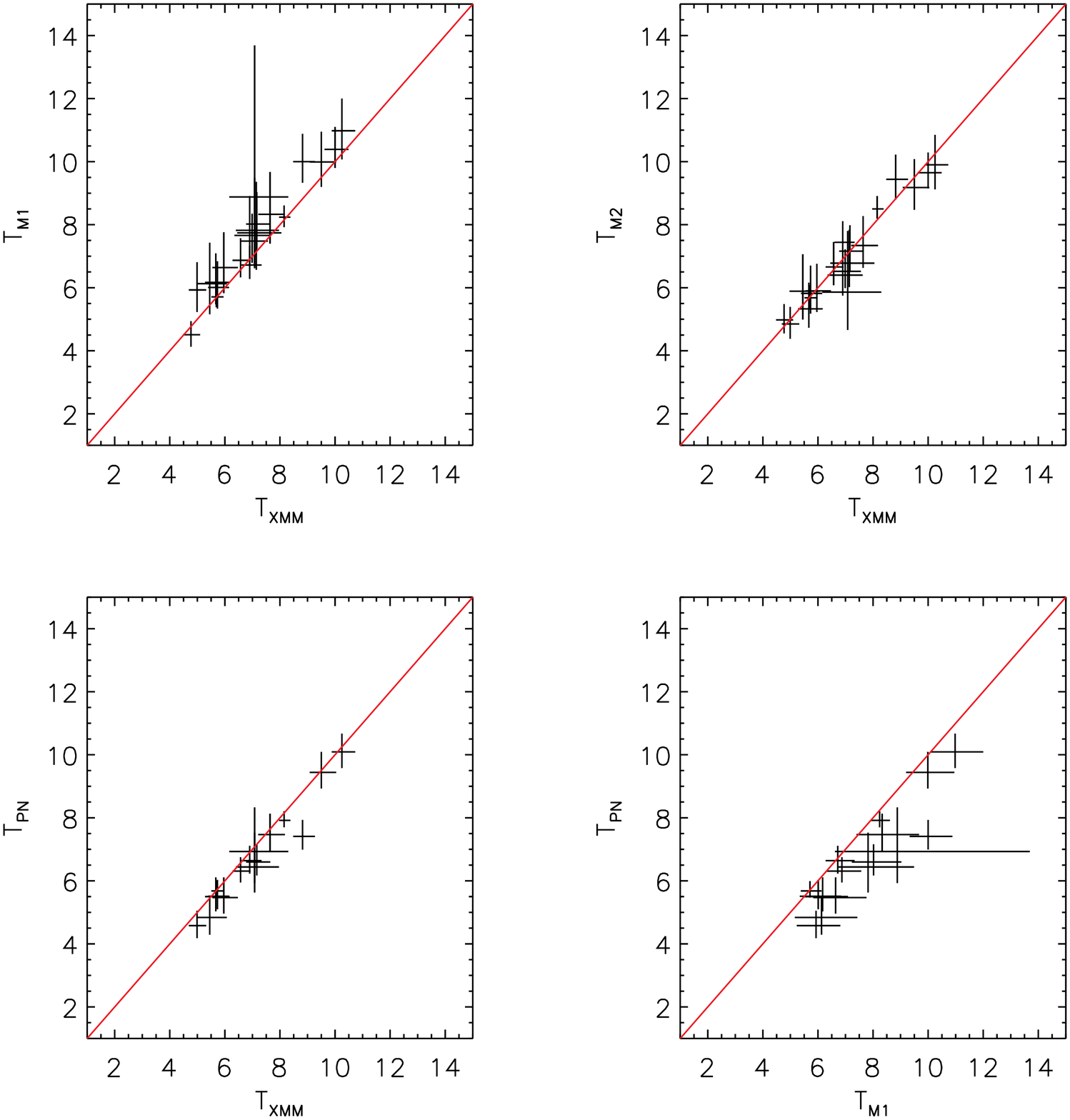}
    \caption{Intercomparison of temperatures derived from the
      different cameras of \emph{XMM-Newton}. The solid red line is the 1:1
      relation.  \textit{Top left panel}: \emph{XMM-Newton} MOS1 temperatures
      versus the temperatures averaged over all three \emph{XMM-Newton}
      cameras.  \textit{Top right panel}: MOS2 versus \emph{XMM-Newton}
      average.  \textit{Bottom left panel}: PN versus \emph{XMM-Newton}
      average.  \textit{Bottom right panel}: PN versus MOS1.}
    \label{fig:xmm_xmm_temp}
  \end{center}
\end{figure*}

To quantify the average effect of the above calibration issues on the
temperature measurements, we fit in the energy band [0.7-10] KeV an
absorbed APEC model to the spectra extracted from each single camera
(ACIS-I, MOS1, MOS2 and PN) and for the combination of the three
cameras of \emph{XMM-Newton} (Table \ref{tab:T_cxo_xmm}).  In Figure~\ref{fig:cxo_xmm_temp} we compare the
temperatures from the three cameras of \emph{XMM-Newton} with the one from
\emph{Chandra}.  For convenience in red we overplot the 1:1 relation.  We
notice that, due to the relatively large errors, if we compare the
temperatures of each cluster, those are consistent within the errors
at 1$\sigma$.  Nevertheless, if we consider the sample, we immediately
notice that, on average, \emph{Chandra} returns higher temperatures than
\emph{XMM-Newton}.  In particular, using the method described in \S6.1, we derived the ratio of $T_{XMM}/T_{CXO}$ for the different
cameras as reported in table \ref{tab:Tratio_cxo_xmm}. We notice that
the average discrepancy is of the order of 9$\pm$3\%.

For completeness, we repeat the same analysis using only the cameras
of \emph{XMM-Newton}. This comparison is shown in Figure~\ref{fig:xmm_xmm_temp}
and the results of the fit are reported in table
\ref{tab:Tratio_cxo_xmm}.  It is important to say that we find that
even \emph{XMM-Newton} alone shows significant average temperature
discrepancy for some cameras.  In particular, we notice that
$T_{pn}/T_{M1}=0.92 \pm 0.3$.

 Finally, we investigate the possibility of a systematic trend with cluster
  temperature by fitting a two-parameter model to
  the respective temperature measurements: $T_1=A\,T_2^\alpha$, where
  $T_1$ and $T_2$ are the temperatures obtained from different
  satellites/cameras, the exponent $\alpha$ characterizes the
  temperature dependence of any bias.  We obtain best fits of
  $\alpha=0.81\pm 0.18, 0.92\pm 0.18, 0.83 \pm 0.14$ for the  PN/ACIS-I, MOS1/ACIS-I,
    MOS2/ACIS-I combinations respectively, indicating that the differences between satellites and instruments  depend on cluster temperature.

\begin{table*}
  \begin{center}
    \caption[]{Comparison of \emph{Chandra} ACIS-I and \emph{XMM-Newton}  temperature estimates\\}
    % \resize
    \begin{tabular}{cccccc}
      \hline
      \noalign{\smallskip}
      Cluster & $T_{Chandra}$ &  $T_{XMM,3cameras}$ & $T_{XMM,MOS1} $ & $T_{XMM,MOS2} $& $T_{XMM,EPN} $ \\ 
      \noalign{\smallskip}
      \hline
           ABELL\,0115s   &$  6.46  ^{+  0.51   }_{-  0.51      }$   &$  5.69  ^{+  0.26   }_{-  0.26      }$   &$  5.71  ^{+  0.42   }_{-  0.42      }$   &$  5.68  ^{+  0.40   }_{-  0.40      }$   &$  5.68  ^{+  0.31   }_{-  0.31    }$\\
\noalign{\smallskip}
          ABELL\,1689   &$ 11.23  ^{+  1.06   }_{-  1.06      }$   &$  9.50  ^{+  0.54   }_{-  0.54      }$   &$  9.99  ^{+  0.96   }_{-  0.96      }$   &$  9.18  ^{+  0.90   }_{-  0.90      }$   &$  9.44  ^{+  0.65   }_{-  0.65    }$\\
\noalign{\smallskip}
          ABELL\,1763   &$  7.98  ^{+  1.45   }_{-  1.45      }$   &$  6.90  ^{+  0.72   }_{-  0.72      }$   &$  7.66  ^{+  1.25   }_{-  1.25      }$   &$  6.40  ^{+  0.80   }_{-  0.80      }$   &$  0.00  ^{+  0.00   }_{-  0.00    }$\\
\noalign{\smallskip}
          ABELL\,1835   &$ 11.06  ^{+  1.09   }_{-  1.09      }$   &$ 10.00  ^{+  0.49   }_{-  0.49      }$   &$ 10.39  ^{+  0.71   }_{-  0.71      }$   &$  9.65  ^{+  0.64   }_{-  0.64      }$   &$  0.00  ^{+  0.00   }_{-  0.00    }$\\
\noalign{\smallskip}
          ABELL\,1914   &$  8.57  ^{+  1.57   }_{-  1.57      }$   &$  7.16  ^{+  0.49   }_{-  0.49      }$   &$  8.02  ^{+  1.01   }_{-  1.01      }$   &$  7.16  ^{+  0.82   }_{-  0.82      }$   &$  6.60  ^{+  0.56   }_{-  0.56    }$\\
\noalign{\smallskip}
           ABELL\,0209   &$  7.56  ^{+  1.40   }_{-  1.40      }$   &$  7.64  ^{+  0.54   }_{-  0.54      }$   &$  8.33  ^{+  1.34   }_{-  1.34      }$   &$  7.34  ^{+  0.93   }_{-  0.93      }$   &$  7.47  ^{+  0.66   }_{-  0.66    }$\\
\noalign{\smallskip}
          ABELL\,2204   &$ 10.50  ^{+  1.11   }_{-  1.11      }$   &$  8.82  ^{+  0.45   }_{-  0.45      }$   &$ 10.00  ^{+  0.88   }_{-  0.88      }$   &$  9.44  ^{+  0.78   }_{-  0.78      }$   &$  7.41  ^{+  0.52   }_{-  0.52    }$\\
\noalign{\smallskip}
          ABELL\,2537   &$  8.68  ^{+  3.78   }_{-  3.78      }$   &$  7.08  ^{+  1.22   }_{-  1.22      }$   &$  8.88  ^{+  4.81   }_{-  4.81      }$   &$  5.86  ^{+  1.94   }_{-  1.94      }$   &$  6.93  ^{+  1.40   }_{-  1.40    }$\\
\noalign{\smallskip}
          ABELL\,2631   &$  6.50  ^{+  1.20   }_{-  1.20      }$   &$  7.08  ^{+  0.88   }_{-  0.88      }$   &$  7.82  ^{+  1.67   }_{-  1.67      }$   &$  0.00  ^{+  0.00   }_{-  0.00      }$   &$  6.44  ^{+  1.09   }_{-  1.09    }$\\
\noalign{\smallskip}
          ABELL\,2813   &$  5.48  ^{+  1.13   }_{-  1.13      }$   &$  5.45  ^{+  0.62   }_{-  0.62      }$   &$  6.13  ^{+  1.30   }_{-  1.30      }$   &$  5.89  ^{+  1.17   }_{-  1.17      }$   &$  4.84  ^{+  0.67   }_{-  0.67    }$\\
\noalign{\smallskip}
           ABELL\,0383   &$  5.76  ^{+  1.26   }_{-  1.26      }$   &$  4.99  ^{+  0.33   }_{-  0.33      }$   &$  5.93  ^{+  0.88   }_{-  0.88      }$   &$  4.85  ^{+  0.54   }_{-  0.54      }$   &$  4.58  ^{+  0.47   }_{-  0.47    }$\\
\noalign{\smallskip}
            ABELL\,0068   &$  5.02  ^{+  1.65   }_{-  1.65      }$   &$  5.67  ^{+  0.50   }_{-  0.50      }$   &$  6.17  ^{+  0.92   }_{-  0.92      }$   &$  5.33  ^{+  0.83   }_{-  0.83      }$   &$  5.50  ^{+  0.61   }_{-  0.61    }$\\
\noalign{\smallskip}
           ABELL\,0773   &$  8.64  ^{+  2.05   }_{-  2.05      }$   &$  7.14  ^{+  0.91   }_{-  0.91      }$   &$  7.74  ^{+  1.62   }_{-  1.62      }$   &$  6.78  ^{+  1.06   }_{-  1.06      }$   &$  0.00  ^{+  0.00   }_{-  0.00    }$\\
\noalign{\smallskip}
           ABELL\,0781   &$  5.64  ^{+  2.22   }_{-  2.22      }$   &$  6.99  ^{+  0.57   }_{-  0.57      }$   &$  7.48  ^{+  0.87   }_{-  0.87      }$   &$  6.52  ^{+  0.70   }_{-  0.70      }$   &$  0.00  ^{+  0.00   }_{-  0.00    }$\\
\noalign{\smallskip}
          RXC\,J1504.1$-$0248    &$ 11.98  ^{+  4.23   }_{-  4.23      }$   &$ 10.25  ^{+  0.48   }_{-  0.48      }$   &$ 10.98  ^{+  1.02   }_{-  1.02      }$   &$  9.90  ^{+  0.95   }_{-  0.95      }$   &$ 10.09  ^{+  0.58   }_{-  0.58    }$\\
\noalign{\smallskip}
          RX\,J1720.1$+$2638   &$  7.46  ^{+  1.03   }_{-  1.03      }$   &$  6.90  ^{+  0.43   }_{-  0.43      }$   &$  6.72  ^{+  0.61   }_{-  0.61      }$   &$  7.44  ^{+  0.67   }_{-  0.67      }$   &$  6.64  ^{+  0.47   }_{-  0.47    }$\\
\noalign{\smallskip}
          RX\,J2129.6$+$0005   &$  7.62  ^{+  1.35   }_{-  1.35      }$   &$  6.57  ^{+  0.36   }_{-  0.36      }$   &$  6.87  ^{+  0.70   }_{-  0.70      }$   &$  6.66  ^{+  0.79   }_{-  0.79      }$   &$  6.31  ^{+  0.44   }_{-  0.44    }$\\
\noalign{\smallskip}
          ZwCl1459.4$+$4240    &$  6.41  ^{+  2.76   }_{-  2.76      }$   &$  5.73  ^{+  0.42   }_{-  0.42      }$   &$  6.01  ^{+  0.83   }_{-  0.83      }$   &$  5.82  ^{+  0.88   }_{-  0.88      }$   &$  5.51  ^{+  0.52   }_{-  0.52    }$\\
\noalign{\smallskip}
          ZwCl\,1021.0$+$0426  &$ 10.48  ^{+  2.10   }_{-  2.10      }$   &$  8.15  ^{+  0.23   }_{-  0.23      }$   &$  8.24  ^{+  0.37   }_{-  0.37      }$   &$  8.50  ^{+  0.41   }_{-  0.41      }$   &$  7.92  ^{+  0.29   }_{-  0.29    }$\\
\noalign{\smallskip}
          ZwCl\,1454.8$+$2233   &$  4.95  ^{+  0.48   }_{-  0.48      }$   &$  4.77  ^{+  0.33   }_{-  0.33      }$   &$  4.51  ^{+  0.43   }_{-  0.43      }$   &$  4.98  ^{+  0.50   }_{-  0.50      }$   &$  0.00  ^{+  0.00   }_{-  0.00    }$\\
\noalign{\smallskip}
           ABELL\,0907   &$  6.23  ^{+  0.45   }_{-  0.45      }$   &$  5.96  ^{+  0.51   }_{-  0.51      }$   &$  6.64  ^{+  1.12   }_{-  1.12      }$   &$  5.90  ^{+  0.86   }_{-  0.86      }$   &$  5.47  ^{+  0.64   }_{-  0.64    }$\\
\noalign{\smallskip}             
            \hline
    \end{tabular}
    \label{tab:T_cxo_xmm}
  \end{center}
\end{table*}

\begin{table*}
  \begin{center}
    \caption[]{Comparison of \emph{Chandra} ACIS-I and \emph{XMM-Newton}  temperature estimates\\}
    % \resize
    \begin{tabular}{l|cccc}
      \hline
      &  $T_{XMM,3cameras}$ & $T_{XMM,MOS1} $ & $T_{XMM,MOS2} $& $T_{XMM,EPN} $ \\ 
      \hline
      $T_{CXO}$& 0.91 $\pm$ 0.03 & 0.95$\pm$ 0.04 & 0.90 $\pm$ 0.04 & 0.85 $\pm$ 0.03 \\
      $T_{XMM,3cameras}$   & -  & 1.04 $\pm$ 0.02 & 1.00 $\pm$ 0.02 & 0.96 $\pm$ 0.02\\
      $T_{XMM,MOS2}$   & -  &  1.04$\pm$0.03&  -  & 0.94 $\pm 0.03$\\
      $T_{XMM,MOS1}$   & -  &  - &  -  & 0.92 $\pm 0.03$\\
      \hline
    \end{tabular}
    \label{tab:Tratio_cxo_xmm}
  \end{center}
\end{table*}

\newpage
\bibliography{ms}

\end{document}